\documentclass[twocolumn,epjc3]{svjour3}          

\RequirePackage[T1]{fontenc}

\RequirePackage{graphicx}
\RequirePackage{mathptmx}      
\RequirePackage{flushend}
\RequirePackage[numbers,sort&compress]{natbib}
\RequirePackage[colorlinks,citecolor=blue,urlcolor=blue,linkcolor=blue]{hyperref}

\usepackage{epsf} 
\usepackage{amsfonts}
\usepackage{amsmath}
\usepackage{amssymb}
\usepackage{mathtools}
\usepackage{bm}
\usepackage{float}
\usepackage{slashed}
\usepackage{dcolumn}
\usepackage{multirow}
\usepackage{color}
\usepackage[caption=false]{subfig} 
\usepackage{cancel} 

\usepackage{xcolor}
\usepackage{soul}

\def\dd{{\mathrm{d}}}

\mathchardef\-="2D

\journalname{Eur. Phys. J. C}

\begin{document}

\title{Comparison of two Minkowski-space approaches to heavy quarkonia}

\author{Sofia Leit\~ao \thanksref{e1,addr1}
        \and
        Yang Li\thanksref{addr2}
        \and 
        Pieter Maris\thanksref{addr2}
        \and 
        M.T. Pe\~na\thanksref{addr3,addr1}
         \and 
        Alfred Stadler\thanksref{addr4,addr1}
        \and
        James P. Vary\thanksref{addr2}
	\and 
	Elmar P.~Biernat\thanksref{addr1}
        }

\thankstext{e1}{e-mail: sofia.leitao@tecnico.ulisboa.pt}

\institute{CFTP, Instituto Superior T\'ecnico, Universidade de Lisboa, Av.\ Rovisco Pais 1, 1049-001 Lisboa, Portugal \label{addr1}
          \and
          Department of Physics and Astronomy, Iowa State University, Ames, IA 50011, USA \label{addr2}
          \and 
          Departamento de F\'isica, Instituto Superior T\'ecnico, Universidade de Lisboa, Av.\ Rovisco Pais 1, 1049-001 Lisboa, Portugal \label{addr3}
          \and
          Departamento de F\'isica, Universidade de \'Evora, 7000-671 \'Evora, Portugal  \label{addr4}
}

\date{Received: date / Accepted: date}

\maketitle

\begin{abstract}
 In this work we compare mass spectra and decay constants obtained from two recent, independent, and
fully relativistic approaches to the quarkonium bound-state problem: the Hamiltonian basis light-front quantization (BLFQ) approach, where
light-front wave functions are naturally formulated; and, the covariant spectator theory (CST),
based on a reorganization of the Bethe-Salpeter equation. Even though {\it conceptually} different, both solutions are obtained in Minkowski space.  Comparisons of decay constants for more than ten states of charmonium and bottomonium show favorable agreement between the two approaches as well as with experiment where available.  We also apply the Brodsky-Huang-Lepage
 prescription
to convert the CST amplitudes into functions of light-front variables. This provides an ideal opportunity to investigate the
similarities and differences at the level of the wave functions. Several qualitative features are observed in remarkable agreement between the two approaches even for the
rarely addressed excited states. Leading twist distribution amplitudes as well as parton distribution functions of heavy quarkonia are also analyzed.
\end{abstract}

\section{Introduction}
\label{sec:intro}

In the last decade the renaissance of interest in quarkonium systems has been driven by the discovery of new particles such as the $X(3872)$ state. Ever since, quarkonium spectroscopy has enjoyed an intensive flow of new results provided by B- and charm-factories such as Belle, BaBar and BES III. Also, several experiments at hadron machines such as the LHC can now investigate quarkonium produced promptly in high-energy hadronic collisions, in addition to charmonium produced in B-decays. Data samples with unprecedented statistics are now available and additional data are anticipated with the advent of  SuperKEKB, the new B factory at KEK. (A detailed review on the experimental status can be seen in \cite{Brambilla} and references therein).

Theoretically, this situation represents an exciting opportunity where QCD inspired models in parallel with lattice calculations can be extensively tested. In this line of investigation, two recent models of QCD, the covariant spectator theory (CST) and the Hamiltonian basis light-front quantization (BLFQ) obtained a successful description of heavy quarkonia below open flavor thresholds.

 In this work we apply improved versions of the models explored in Refs.\,\cite{Leitao:2016bqq} for CST and \cite{Li:2015zda} for BLFQ respectively, and extend their range of results with new sets of predictions for mass spectra, decay constants and light-front distributions. We find that both approaches consistently succeed in describing the experimental data, despite the limitations intrinsic to each model.

In CST, a quasi-potential equation is obtained by reorganizing the Bethe-Salpeter equation (BSE) and solving for a given kernel. On the other hand, in BLFQ, solutions are obtained by diagonalizing an effective QCD Hamiltonian. While the goals of both approaches are the same - to formulate a successful relativistic model of the mesons in terms of quark-antiquark degrees of freedom - they are distinct, especially in the way certain features of QCD are implemented. This motivates a detailed comparison of these approaches as well as the resulting observables.  We focus on heavy quarkonia where we have the advantage of making useful comparisons at the non-relativistic limit.

We emphasize at the outset that in both approaches quarkonium is treated non-perturbatively as a relativistic bound-state. Both approaches include a one-gluon exchange interaction, known to be essential for a proper description of the low-lying heavy quarkonia, as well as a confining interaction.

Another feature that both models possess is the fact that they are formulated {\it directly} in Minkowski space-time, making them complementary to other approaches such as Euclidean Dyson--Schwinger Equations (DSE) \cite{Bhagwat,Blank,Hilger,Popovici,DSE2016} and Lattice QCD \cite{McNeile}. In fact, the shared Minkowskian nature of CST and BLFQ invites a detailed comparison. Here we pursue this comparison as follows: first, we take the results for quarkonia of the two approaches and extend them by providing new sets of physical observables obtained with improvements in each approach and presented here in parallel for convenience; secondly, we compare these results and try to address the natural question -- is it possible to quantify their similarities and differences not only at the level of observables but also in terms of quantities of interest for further applications, e.g.\ light-front wave functions?

Though it is well-known that establishing a proper connection between Bethe-Salpeter amplitudes or any of its three-dimensional reductions, with light-front wave functions is a non-trivial problem, attempts to solve or at least bridge the two approaches have been developed  in multiple contexts \cite{Miller2009,Terry2016,Hobbs2017,Kusaka,Sales2000,Karmanov2006,Frederico2012,Carbonell2017,Karmanov2017,GSalme2017}. 

Here, and to answer the previous question, we have adopted the widely used Brodsky-Huang-Lepage prescription \cite{BHL} (cf.
\cite{BragutaPLB,Bodwin}) to express the CST amplitudes in terms of light-cone coordinates and compare these amplitudes to light-front wave functions calculated from BLFQ.

Notwithstanding formal difficulties and caveats, we produce here good agreement between results of the two approaches not only for the
lowest states but also for the higher radial and angular excited states. In addition we show that, contrary to what
intuition might dictate, a simple map (described below), applied to the CST amplitudes allows us to capture all the qualitative features of
{\it genuine} light-front wave functions.

We also calculate decay constants using the mapped CST amplitudes within the Hamiltonian light-front formalism and compare them  with those
calculated directly within the CST approach. The results are consistent, differing roughly by less than 2\% for bottomonium and 10\% for
charmonium.

This good agreement further motivated us to use these mapped CST amplitudes and the BLFQ light-front wave functions for  calculating other
relevant quantities. Most importantly, we highlight the parton distribution amplitudes (PDAs), whose precise knowledge is crucial for the
study of a panoply of processes such as quarkonia production at high-energies \cite{BragutaPLB,BragutaPRD}, $J/ \psi+\eta_c$ pair production
in $e^+e^-$ annihilation \cite{Choi}, $B_c\rightarrow \eta_c$ transitions \cite{Zhong}, decays of heavy $S$-wave quarkonia into lighter
vector mesons \cite{Braguta2010}; deeply virtual quarkonia production \cite{Koempel} and Higgs boson decays into quarkonia
\cite{Bodwin2014}. 

We also calculate heavy quarkonia parton distribution functions (PDFs) and provide their moments heretofore less studied in the
literature.

This paper is organized as follows: section \ref{sec:formalism} introduces the notation, definitions  and the general formalism employed in
this work; section \ref{sec:results} is devoted to results and discussion and finally, in section \ref{sec:conclusions}, we present a brief
summary, our conclusions and outlook for further research.

\section{Formalism}
\label{sec:formalism}
In this section we introduce the two models used to study heavy quarkonium that we compare in this work.  Motivations,
goals and specific details of each approach can be found correspondingly for CST and BLFQ in Refs.\,\cite{Leitao:2016bqq} and
\cite{Li:2015zda} and the references therein. Here we aim for a brief self-contained description, accompanied by the definitions of
the relevant quantities under discussion in the following sections: decay constants, CST amplitudes, light-front wave functions (LFWFs),
parton distribution amplitudes (PDAs) and parton distribution functions (PDFs).

\subsection{Hamiltonian basis light-front quantization for quarkonium}
In the light-front Hamiltonian approach, quarkonium is described by state
vectors $|\psi_h\rangle$. These state vectors can be obtained by diagonalizing the light-front quantized Hamiltonian
operator $\hat{P}^-$,
\begin{equation}
 \hat P^-|\psi_h(P, J, m_J)\rangle = \frac{\bm{P}^2_\perp+M^2_h}{P^+}|\psi_h(P, J, m_J)\rangle,
 \label{eq:state}
\end{equation}
where $P=(P^-, P^+, \bm{P}_\perp)$ is the 4-momentum; $J$ and $m_J$ are the total angular momentum and the magnetic projection,
respectively. For a 4-vector $v$, the light-front variables are defined as: $v^\pm = v^0\pm v^3$, $\bm{v}_\perp = (v^1, v^2)$.
The eigenvalue equation is usually rewritten as,
\begin{equation}
 (P^+\hat P^--\bm{P}^2_\perp) |\psi_h\rangle = M^2_h |\psi_h\rangle,
\end{equation}
and so it is easy to identify $P^+ P^--\bm{P}^2_\perp = P_\mu P^\mu \equiv H_\textsc{lc}$ as the invariant mass squared
operator, also known as the ``light-cone Hamiltonian''. 

In the BLFQ approach {\cite{Li:2015zda}}, an effective Hamiltonian is adopted based on light-front holographic QCD {\cite{Brodsky:2014yha}}:
\begin{equation}
H_{\text{eff}}\equiv\frac{\bm{k}^2_\perp+m^2}{x(1-x)}+V_{T}+V_L(x)+V_g,
\label{eq:Heff1}
\end{equation}
where $m$ is the mass of the quark; $x=p^+_1/P^+$ is the longitudinal momentum fraction,
\begin{equation}
V_T\equiv \kappa^4 \bm{\zeta}^2_\perp=\kappa^4 x(1-x) \bm{r}_\perp, \label{eq:VLperp}
\end{equation}
is the  ``soft-wall'' light-front holography \cite{Brodsky:2014yha} in the transverse
direction. $\bm{r}_\perp=\bm{r}_{1\perp}-\bm{r}_{2\perp}$ is the transverse separation of the quark and the antiquark, while
\begin{equation}
V_L \equiv -\frac{\kappa^4}{4m^2}\partial_x(x(1-x)\partial_x),
 \label{eq:VLx}
\end{equation}
is the longitudinal confining potential introduced for the first time in Ref.\,\cite{Li:2015zda}. The partial derivative $\partial_x$ is
taken with respect to the holographic variable $\bm \zeta_\perp$, viz $\partial_x f(x,\bm \zeta_\perp)|_{\bm \zeta_\perp}$. The strength
of both $V_T$ and $V_L$ depends on a confinement parameter $\kappa$.
\begin{equation}
V_g\equiv -\frac{C_F 4\pi \alpha_s(Q^2)}{Q^2}\bar{u}_{s'}(k'_1)\gamma_\mu u_s(k_1)\bar{v}_{\bar{s}}(k_2)\gamma^\mu v_{\bar{s}'} (k'_2),
\label{eq:Vg}
\end{equation}
is the one-gluon exchange term with
\begin{align}
Q^2=-\bar{q}^2=-(1/2)(k'_1-k_1)^2-(1/2)(k'_2-k_2)^2,
\end{align}
the average four-momentum of the exchanged gluon and $C_F=4/3$. As an extension to the work presented in \cite{Li:2015zda},
instead of using a fixed value for $\alpha_s$, a running coupling is used here (see Ref.~\cite{Li:2016} for
details).

\subsection{Light-front wave functions (LFWFs)}
Light-front wave functions (LFWFs) are defined from the Fock space expansion of the state vector in Eq.~(\ref{eq:state}). For
example, the quarkonium LFWF within the valence sector reads,
\begin{multline}\label{eqn:Fock_expansion}
 |\psi_h(P, J, m_J)\rangle = 
\sum_{s, \bar s}\int_0^1\frac{\dd x}{2x(1-x)} \int \frac{\dd^2 {\bm k}_\perp}{(2\pi)^3}\\
\times \psi^{(m_J)}_{s\bar s/h}(\bm{k}_\perp, x) \frac{1}{\sqrt{N_c}}\sum_{i=1}^{N_c} b^\dagger_{s{}i}(xP^+, \bm{k}_\perp+x \bm{P}_\perp) \\
\times d^\dagger_{\bar s{}i}\big((1-x)P^+, -\bm{k}_\perp+(1-x)\bm{P}_\perp\big) |0\rangle. 
\end{multline}
Here $N_c=3$ is the number of colors, and $\bm{k}_\perp \equiv \bm{p}_\perp - x \bm{P}_\perp$ is the relative transverse momentum. 
The quark and antiquark creation operators $b^\dagger$ and $d^\dagger$ satisfy the canonical anti-commutation relations,
\begin{equation}\label{eqn:CCR}
 \begin{split}
& \big\{ b_{si}(p^+, \bm{p}_\perp), b_{s'i'}^\dagger(p'^+, \bm{p}'_\perp) \big\}   \\
=\,& \big\{ d_{si}(p^+,\bm{p}_\perp), d_{s'i'}^\dagger(p'^+, \bm{p}'_\perp) \big\}  \\
=\,& 2p^+(2\pi)^3\delta(p^+-p'^+)\delta^2(\bm p_\perp-\bm p'_\perp)\delta_{ss'}\delta_{ii'}\,. \\
 \end{split}
\end{equation}
The state vector is normalized according to a one-particle state [cf. Eq.~(\ref{eqn:CCR})]:
\begin{multline}\label{eqn:normalizaiton_of_state_vector}
 \langle \psi_h(P, J, m_J) | \psi_{h'}(P', J', m_J')\rangle =  
 2P^+ (2\pi)^3 \\
\times \delta(P^+ - P'^+)\delta^2(\bm P_\perp - \bm P'_\perp)
\delta_{JJ'}\delta_{m_J,m_J'} \delta_{hh'}.
\end{multline}
Then, the normalization of the LFWFs reads, 
\begin{multline}\label{eqn:normalization}
 \sum_{s, \bar s} \int_0^1\frac{\dd x}{2x(1-x)} \int \frac{\dd^2  {\bm k}_\perp}{(2\pi)^3} 
\psi^{(m_J')*}_{s \bar s/h'}(\bm{k}_\perp, x)\psi^{(m_J)}_{s \bar s/h}(\bm{k}_\perp, x)  \\
= \delta_{hh'}\delta_{m_J,m_J'}.
\end{multline}
Note that the state vectors of different particles, e.g., $J/\psi$ and $\psi'$, are orthogonal.

\subsection{Covariant spectator theory for quarkonium}
\label{sec:CST}
In CST, quarkonium is described as a relativistic system of a quark and antiquark, bound together by a QCD-inspired interaction. The CST equation can be derived from the BSE. For a bound state of total
four-momentum $P$ coupled to a quark with momentum $p_1=p+\frac{1}{2}P$ and an antiquark with momentum $-p_2=-p+\frac{1}{2}P$, the BSE reads
\begin{multline}
\Gamma_{\text{BS}}(p_1,p_2)= i\int\frac{ {\dd}^4k}{(2\pi)^4}\,{\cal V}(p,k;P) 
\\
\times S({k}_1)\,\Gamma_{\text{BS}}(k_1,k_2)\,S(k_2)\,,
\label{eq:BS}
\end{multline}
where $S(k_i)$ is in principle the dressed quark propagator. However, in this work, $S$ has been replaced by the bare propagator. The idea of CST is to approximate this equation by keeping in the $k_0$-contour integration {\it only} the contribution from the positive-energy pole of one quark
propagator (for details on this prescription see \cite{Gro69}). This leads to the so called
one-channel spectator equation (1CSE), given by
\begin{multline}
\Gamma_{\text{1CS}} (\hat{p}_1,p_2)= - \int \frac{\dd^3 {\bm k}}{(2\pi)^3} \frac{m}{E_{k}} \sum_K V_K(\hat{p}_1,\hat{k}_1) \Theta_{1}^{K(\mu)}  
\\
\times
\frac{m+\hat{\slashed{k}}_1}{m} \Gamma_{\text{1CS}}(\hat{k}_1,k_2)
\frac{m+\slashed{k}_2}{m^2-k_2^2-i\epsilon}\Theta^K_{2(\mu)} \, ,
\label{eq:1CS}
\end{multline}
where $\Theta_i^{K(\mu)}=\mathbf{1}_i, \gamma^5_i,$ or $\gamma_i^\mu$;  the functions $V_K(\hat{p}_1,\hat{k}_1)$ describe the momentum dependence of the kernel, $m$ is the constituent mass of the quarks, and 
\begin{equation}
E_{k}\equiv (m^2+\bm{k}^2)^{1/2}.
\label{eq:energy}
\end{equation}
Note that in context of CST a ``$\hat{\phantom{p}}$'' over a four-momentum indicates that the particle is on-mass-shell, and that we use ${\bm k}$ to indicate 3-momenta.  It is worth mentioning that
Eq.~(\ref{eq:1CS}) retains from the BSE four important properties: manifest covariance, cluster separability, and the correct one-body and nonrelativistic limits.

The kernel we employed in the 1CSE consists of a covariant generalization of the linear (lin) confining potential used in
Ref.~\cite{Leitao:2014}, a one-gluon exchange (OGE), and a constant (C) interaction:
\begin{multline}
\sum_K V_K  \Theta_{1}^{K(\mu)} \otimes \Theta^K_{2(\mu)}\,= 
\\
\left[ (1-y) \left({\bf 1}_1\otimes {\bf 1}_2 + \gamma^5_1 \otimes \gamma^5_2 \right) - y\, \gamma^\mu_1 \otimes \gamma_{\mu 2} \right]V_\mathrm{lin}  -
\\
\gamma^\mu_1 \otimes \gamma_{\mu 2} \left[ V_\mathrm{OGE}+V_\mathrm{C} \right]\,.
\label{eq:kernel}
\end{multline}
The Lorentz structure of the confining kernel is flexible: the mixing parameter $y$  allows one to dial between a scalar-plus-pseudoscalar structure, which preserves chiral symmetry as shown
in Ref.~\cite{CSTpi-pi}, and a vector structure, while leaving the nonrelativistic limit unchanged.

 An analysis of the asymptotic behavior or large momenta  $|{\bf k}|$ shows that we need to regularize the kernel. We use Pauli-Villars regularization for both the linear and the OGE parts, which yields one additional parameter, the cut-off parameter $\Lambda$. We found that $\Lambda=2 m$ is a reasonable choice. 
 
 The momentum-dependent
terms of the interaction kernel are
\begin{multline}
V_\mathrm{lin}(p,k)  = -8\sigma \pi\left[\left(\frac{1}{q^4}-\frac{1}{\Lambda^4+q^4}\right)-
\right. \\
    \left. 
\frac{E_{p}}{m}(2\pi)^3 \delta^3 (\bm{q})\int \frac{\dd^3
{\bm k}'}{(2\pi)^3}\frac{m}{E_{k'}}\left(\frac{1}{q'^4}-\frac{1}{\Lambda^4+q'^4}\right)\right],
\end{multline}

\begin{align}
&V_\mathrm{OGE}(p,k)  = -4 \pi \alpha_s \left(\frac{1}{q^2}-\frac{1}{q^2-\Lambda^2}\right), \\
&V_\mathrm{C}(p,k) = (2\pi)^3\frac{E_{k}}{m} C \delta^3 (\bm{q})\,,
\label{eq:V}
\end{align}
where $q^{(\prime)}=p-k^{(\prime)}$.

As an extension to the model described in Ref.~\cite{Leitao:2016bqq}, we now allow the constituent masses to be free parameters. We also let
 the parameter $y$ to be determined by the fit,  in order to examine how much the Lorentz structure of the kernel is constrained by the mass spectra. 
These parameters are fitted exclusively and simultaneously to $b\bar{b}$ and $c\bar{c}$ states.

\subsection{CST amplitudes}
\label{sec:CSTamp}
The solutions of Eq.~(\ref{eq:1CS}) have been determined in the meson rest-frame, where $P=(M, \bm{0})$ and $M$ is the meson mass. Furthermore,
instead of solving for $\Gamma_{\text{1CS}}$ directly, we solved for CST amplitudes, $\Psi^{+\rho}_{\lambda_1\lambda_2}$,
 defined as 
\begin{equation}
 \Psi^{+\rho}_{\lambda_1\lambda_2}(\bm{k}) \equiv \frac{m}{E_k}\frac{\rho}{(1-\rho)E_{k}+\rho M}
\Gamma^{+\rho}_{\lambda_1\lambda_2}(\bm{k}),
\label{eq:CSTampDef}
\end{equation}
where $\Gamma^{+\rho}_{\lambda_1\lambda_2} (\bm{k})  \equiv \bar{u}_1^+(\bm{k},\lambda_1) \Gamma(k) u_2^{\rho}(\bm{k},\lambda_2)$ and
$u^\rho$ with $\rho=\pm$ are helicity $\rho$-spinors, as given in Ref.~\cite{SLquarkconf}.

These amplitudes can be expanded in a very useful basis,
\begin{equation}
\Psi^{+\rho}_{\lambda_1\lambda_2}(\bm{k}) =\sum_j \psi_j^\rho(|\bm{k}|) \chi^\dagger_{\lambda_1}(\hat{\bm{k}})\, K_j^\rho(\hat{\bm k}) \,
\chi_{\lambda_2}(\hat{\bm k}),
\label{eq:CSTamplitudes}
\end{equation}
where $\hat{\bm k}\equiv {\bm k}/|{\bm k}|$, $\chi_\lambda$ are two-component helicity spinors and the $K_j^\rho(\hat{\bm k})$ operators are
$2\times 2$ matrices that depend on the total angular momentum $J$ and the parity $P$ of the meson under study. A list with all
$K_j^\rho(\hat{\bm k})$ operators used in this work can be seen in Ref.~\cite{SLquarkconf}. 

In terms of these scalar wave functions, the pseudoscalar and vector mesons are normalized according to 
\begin{equation}
1=\frac{N_c}{4M\pi^2} \int \dd |{\bm k}| \,{\bm k}^2 \left(\psi^2_{s}(|{\bm k}|)+\psi^2_{p}(|{\bm k}|)\right).
\label{eq:normPS_S}
\end{equation}
Similarly, for vector and axial vector mesons the normalization condition is
\begin{multline}
1=\frac{N_c}{4M\pi^2} \int \dd |{\bm k}| \, {\bm k}^2 \left(\psi^2_{s}(|{\bm k}|)+\psi^2_{d}(|{\bm k}|)+
\right. \\ \left.
\psi^2_{p_s}(|{\bm k}|)+\psi^2_{p_t}(|{\bm k}|)\right),
\label{eq:normV_AV}
\end{multline}
where $M$ is the mass of the bound-state, $N_c$ is the number of colors and $\psi_s,$ $\psi_d$, $\psi_{p_s}$ and $\psi_{p_t}$ refer to $S$-, $D$-, singlet and triplet P-waves, respectively.

Both the total angular momentum $J$ and parity $P$ are exact quantum numbers of the CST solutions.  Expressed as in (\ref{eq:CSTamplitudes}), the CST equation is transformed into a system of coupled partial-wave equations, where each partial wave has a definite orbital angular momentum and total spin. Thus it is straightforward to identify the angular momentum of each state which is useful when comparing with experiment. On the other hand, the 1CSE solutions do not have a definite charge conjugation parity, $C$. However, this can be remedied when the appropriately symmetrized contributions of all four poles of the two quark propagators are taken into account in the $k^0$ contour integration, and a coupled four-channel equation is solved instead (cf. discussion in Ref.~\cite{Leitao:2016bqq}).

\subsection{Brodsky-Huang-Lepage prescription}
\label{sec:BHLmap}
Having specified our models, we now describe our method for converting the CST amplitudes of Eq.~(\ref{eq:CSTamplitudes}) into LFWFs. 

The covariance of the CST equations allows us to evaluate the longitudinal momentum fraction $x$ of the  on-shell  quark in the rest-frame ($x$ is
 an  invariant under longitudinal boosts). From the CST kinematics described in section \ref{sec:CST}, the relative four-momentum explicitly reads $k=(k^0,{\bm
k})=(k^0,{\bm k}_\perp, k^3)$, $k_\perp = |\bm{k}_\perp|$, and
\begin{equation}
P^+=M, \qquad  k_1^+=E_k+k^3,
\end{equation}
where $E_k$ is the on-shell energy (\ref{eq:energy}). In the rest-frame the $x$ variable should in principle be identified as \cite{FGross}
\begin{equation}
x=\frac{k_1^+}{P^+}=\frac{E_k+k^3}{M}=\frac{\sqrt{m^2+{\bm k}_{\perp}^2+(k^3)^2}+k^3}{M}.
\label{eq:originalx}
\end{equation}
Consequently,
\begin{equation}
{\bm k}^2=\frac{1}{2}({\bm k}_{\perp}^2 +m^2)+\left(\frac{x M}{2}\right)^2+ \left(\frac{{\bm k}_{\perp}^2 +m^2}{2xM}\right)^2.
\end{equation}
From Eq.~(\ref{eq:originalx}) one verifies that,
\begin{equation}
\displaystyle{\min x =\lim_{k^3 \to -\infty}}x=0, \quad \displaystyle{\max x =\lim_{k^3 \to +\infty}}x=+\infty.
\end{equation}
The last limit poses a difficulty because $x$ can be outside the region $0 \leq x\leq 1.$
How to properly deal with this issue certainly requires further investigation and it is beyond of the scope of this article.
Recent work has been done in that direction, investigating the formal relation between the light-cone and CST box diagrams for a scalar theory \cite{FGross}. In any case, we expect that the contribution to the wave function for
values of $x>1$ should be small, and that it vanishes exactly in the non-relativistic limit \cite{Miller2009}.

We circumvent this difficulty by adopting the Brodsky-Huang-Lepage (BHL) prescription 
\cite{BHL}, where $x$ is automatically limited between 0 and 1, and investigate to what extent such a prescription gives reasonable results.

For the equal mass case of quarkonium $m_q=m_{\bar{q}}=m$, the BHL prescription provides 
 \begin{equation}
 x=\frac{k^+}{P^+}\equiv \frac{E_k+k^3}{2E_k}= \frac{1}{2}+\frac{k^3}{2\sqrt{{k}^2_\perp+(k^3)^2+m^2}}.
 \label{eq:BHLx}
 \end{equation}
From Eq.~(\ref{eq:BHLx})  it is straightforward to derive
\begin{equation}
{\bm k}^2= \frac{{\bm k}_{\perp}^2 +m^2}{4x(1-x)}- m^2.
\label{eq:BHLx2}
\end{equation}

We thus identify the ``CST LFWFs" as:
\begin{equation}
 \psi^{+\rho}_{s_1s_2}(\bm k_\perp, x) \equiv \Psi^{+\rho}_{s_1s_2} \big( \bm k_\perp, k^3(k_\perp, x) \big),
\end{equation}
up to some normalization factors.

\begin{figure}
 \captionsetup[subfloat]{singlelinecheck=off}
\centering
  \subfloat[]{
\includegraphics[width=.24\textwidth]{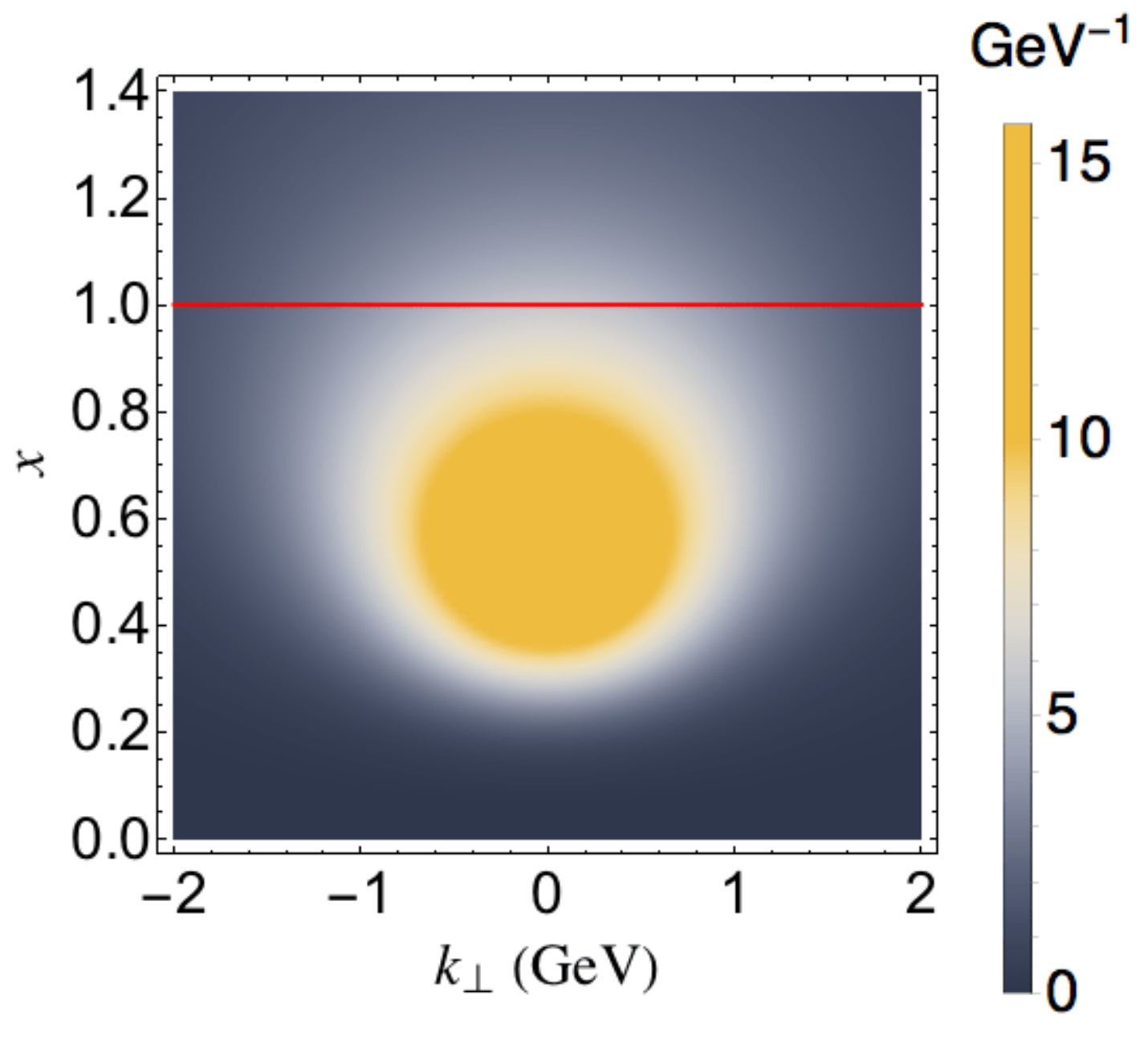}}
  \subfloat[]{%
    \includegraphics[width=.24\textwidth]{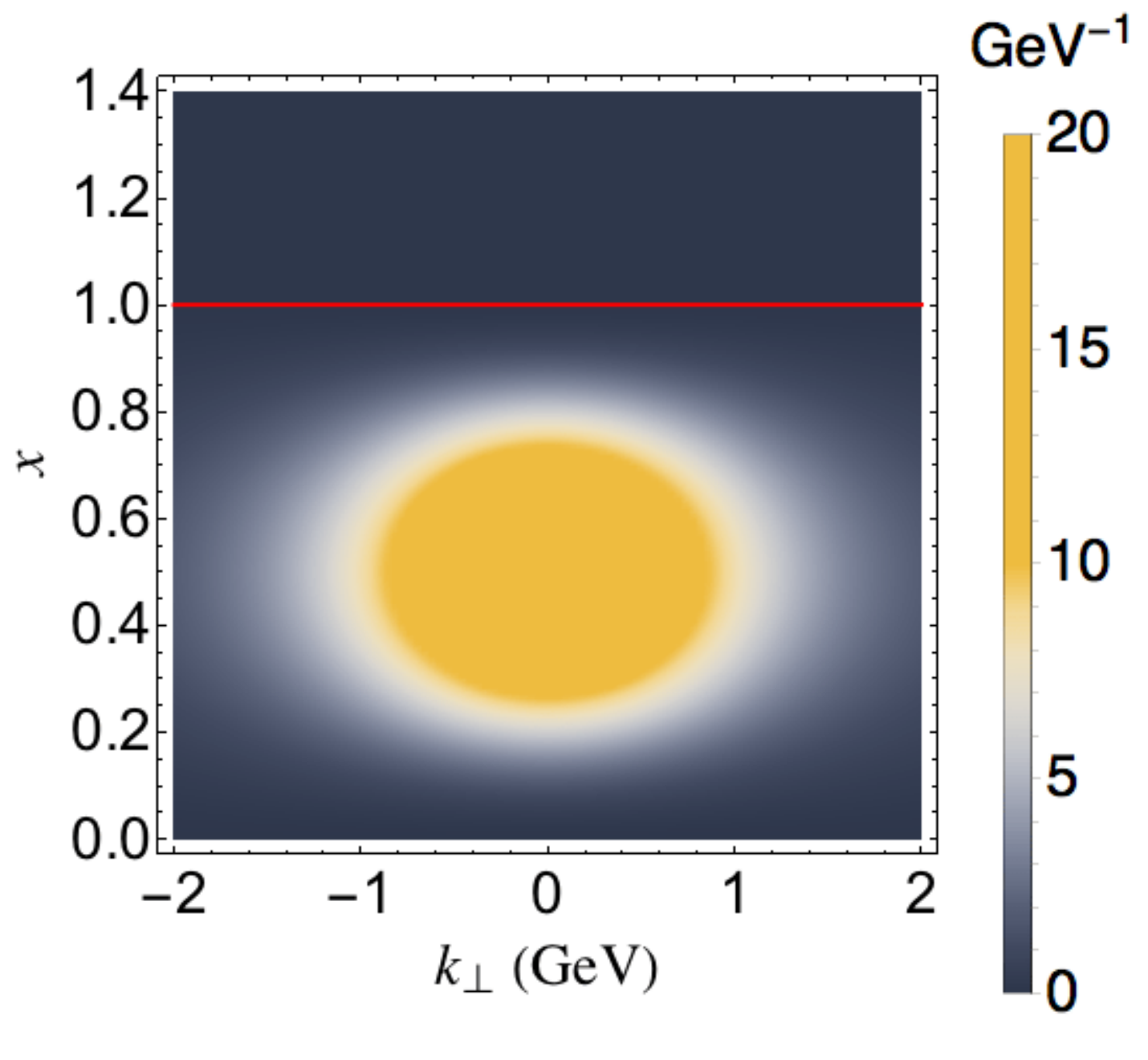}}
  \caption{CST amplitudes for the triplet component of $J/\psi(1S)$ state with $\lambda=0$ using two different changes of coordinates: (a) using the definition of $x$ given in Eq.~(\ref{eq:originalx}) and (b) using the BHL prescription.}
  \label{fig:originalvsBHLmap}
\end{figure}

In Fig.\,(\ref{fig:originalvsBHLmap}) we compare the CST amplitudes for one of  the dominant wave function components of $J/\psi$ after
using the change of variables expressed in (\ref{eq:originalx}) (left panel) {\it vs.}\,the one in (\ref{eq:BHLx}) (right panel). The visualization scheme
is explained in detail in Sec.~\ref{sec:LFWFs}.

The CST amplitude on the left panel spreads beyond the physical region $0\le x \le 1$, but for $x>1$ it is fairly small for the illustrative case of charmonium, where relativity is no longer negligible. On the right panel, the wave function mapped using the BHL prescription is symmetric with respect to $x=0.5$ 
and is restricted to $0\le x \le1$, consistent with longitudinal light-front momentum conservation.

\subsection{Definition of physical observables and distribution functions}
\label{sec:definitions}

Both the LFWFs and the CST amplitudes allow us to calculate a variety of observables. But with the LFWFs obtained from the BHL mapping, one also gains direct access to quantities such as light-cone distributions, whose extraction is not as straightforward in approaches relying on Euclidean formulations.
In this section we apply the LFWFs to the calculation of decay constants and leading-twist parton distribution
amplitudes and parton distribution functions.

\subsubsection{Decay constants}
Decay constants are very important quantities to probe short-range physics. In practice they will be sensitive to the effective short-range
potential. This implies that for any realistic model of quarkonia, having a correct implementation of the one-gluon exchange interaction
is essential for a good description of the decay constants.

In the absence of a proper renormalization procedure, decay constants could develop dependence on the regularization scheme adopted. By construction, any UV regulator estimate within BLFQ is tied to the basis truncation $N_{\text{max}}$. In fact, previous studies indicate that the cut-off
scale is very well approximated by $\Lambda_{UV} \equiv \kappa \sqrt{N_{\text{max}}}$. On the other hand, in CST there is no dependence on
any basis, but $\alpha_s$ has been kept fixed in the CST calculations. Furthermore, in CST the regularization of the integral over $\bm k$ in Eq.~(\ref{eq:1CS}) is governed by the Pauli-Villars cut-off parameter $\Lambda$ and for that reason it will be taken
as the CST estimate for the UV regulator.  Later we will come back to this point when analysing the results obtained for the parton distribution functions. The choice of $\Lambda_{UV} \approx 1.7m$ in BLFQ and
$\Lambda_{UV} \approx 2m$ in CST, permits a good description of the decay constants, with models just fixed by spectroscopy. For the
remainder of the work in BLFQ results this scale cut-off is ensured by choosing $N_{\text{max}}=32$ for bottomonium and $N_{\text{max}}=8$
for charmonium, making the scales of the two the approaches comparable.
\\

In general, the decay constants for pseudoscalar (P), axial-vector (A), {scalar (S)} and vector (V) mesons are defined, respectively, by the matrix elements
\begin{align}
&P^\mu f_P =i \langle 0| \bar{\Psi} \gamma^\mu \gamma^5 \Psi | P\rangle, \label{eq:fP}\\
& \epsilon_\lambda^\mu  m_A f_A=\langle 0| \bar{\Psi} \gamma^\mu \gamma^5 \Psi |A\rangle, \label{eq:fA}\qquad (\lambda=0,\pm1)\\
& P^\mu f_S = \langle 0| \bar{\Psi} \gamma^\mu \Psi | S\rangle, \label{eq:fS}\\
& \epsilon_\lambda^\mu  m_V f_V=\langle 0| \bar{\Psi} \gamma^\mu \Psi |V\rangle, \qquad \quad(\lambda=0,\pm1) \label{eq:fV}
\end{align}
where $P^\mu$ is the total momentum of the meson and the polarization vectors are
\begin{align}
&\epsilon_{\lambda=0}=(0,0,0,1),\\
&\epsilon_{\lambda=\pm1}=\mp\frac{1}{\sqrt{2}}(0,1,\pm i,0).
\end{align}

For pseudoscalar and vector states, BLFQ decay constants are determined as follows,
\begin{align}
f_{P}=&2 \sqrt{N_c} \int_0^1 \frac{\dd x}{2\sqrt{x(1-x)}}\int \frac{\dd^2 {\bm k}_\perp}{(2\pi)^3} \nonumber \\
&\times \left[ \psi_{\uparrow\downarrow}({\bm k}_\perp,x)-\psi_{\downarrow\uparrow}({\bm k}_\perp,x)\right],
\label{eq:BLFQdecayp}\\
f_{V}=&2 \sqrt{N_c} \int_0^1 \frac{\dd x}{2\sqrt{x(1-x)}}\int  \frac{\dd^2 {\bm k}_\perp}{(2\pi)^3} \nonumber \\
&\times\left[ \psi_{\uparrow\downarrow}({\bm k}_\perp,x)+\psi_{\downarrow\uparrow}({\bm
k}_\perp,x)\right], \label{eq:BLFQdecayv}
\end{align} 
and for CST we have 
\begin{equation}
f_P=\int\frac{\dd^3 {\bm k}}{(2\pi)^3}\frac{E_k+m}{M E_{k}}\left[(1-\tilde{k}^2) \psi^P_s(|{\bm k}|)+\tilde{k}^2 \cos\theta \psi^P_p(|{\bm k}|)\right],  \label{eq:CSTdecayp}
\end{equation}
\begin{multline}
f_V=\int\frac{\dd^3 {\bm k}}{(2\pi)^3}\frac{E_k+m}{M E_{k}}\left[(1-\tilde{k}^2)\left(\psi^V_s(|{\bm k}|) +
\right.\right. \\ \left. \left.
    \psi^V_d(|{\bm k}|)(3\cos^2\theta-1)/\sqrt{2}\right)
+\tilde{k}^2 \cos\theta \psi^V_{p_s}(|{\bm k}|)\right],  \label{eq:CSTdecayv}
\end{multline}
where $\tilde{k}=|{\bm k}|/(E_k+m)$ and the scalar functions $\psi(|{\bm k}|)$ are implicitly defined in
(\ref{eq:CSTamplitudes}). 

\subsubsection{Leading-twist parton distribution amplitudes (PDAs)}

In this work we calculate leading-twist parton distribution amplitudes for pseudoscalar $\phi_P(x)$ and  longitudinally polarized vector
$\phi_V^{||}(x)$ mesons. They are determined through the LFWFs as
\begin{multline}
\frac{f_{P,V}}{2\sqrt{2 Nc}}\phi_{P,V^{||}}(x;\mu)=\frac{1}{\sqrt{x(1-x)}}\\
\times
\int\limits_0^{\mathclap{k_\perp \leq \mu}}\frac{\dd^2 {\bm k}_\perp}{2(2\pi)^3}\psi^{\lambda=0}_{\uparrow\downarrow\mp\downarrow\uparrow}({\bm
k}_\perp, x),
\label{eq:DA}
\end{multline}
where $f_{P,V}$ are the previously defined decay constants and $\mu$ is related to the renormalization scale or UV cut-off scale. 
The PDAs defined here satisfy the normalization condition
\begin{equation}
\int_0^1 \phi(x)\dd x=1.
\end{equation}
Moments of these distributions are given by
\begin{equation}
\langle \xi^n  \rangle=\int_0^1  \dd x (2x-1)^n \phi(x).
\end{equation}
In NRQCD, moments are related to the r.m.s.\ relative velocity of the valence quarks \cite{BragutaPLB} through
\begin{equation}
\langle v^n  \rangle=(n+1)\langle \xi^n  \rangle.
\end{equation}

\subsubsection{Parton distribution functions (PDFs)}
Similar to PDAs, parton distribution functions depend on the UV cut-off scale as well.
They can be accessed by
\begin{equation}
f(x;\mu)=\frac{1}{2x(1-x)}\sum_{s,\bar{s}}\int\limits_0^{k_\perp \leq \mu} \frac{\dd^2 {\bm k}_\perp}{(2\pi)^3}|\psi_{s\bar{s}}({\bm
k_\perp}, x)|^2.
\label{eq:PDF}
\end{equation}

From the previous discussion, it is known that both approaches already have built-in regulators and for that reason there is no need for a hard cut-off, so we conveniently extend these integrals to infinity in Eq.~(\ref{eq:DA}) and Eq.~(\ref{eq:PDF}) and drop the reference to $\mu$.

The moments of these distributions are given by
\begin{equation}
\langle x^n  \rangle=\int_0^1  \dd x \,x^n f(x).
\end{equation}

 \section{Results and Discussion}
 \label{sec:results}
We now proceed to the analysis of the results. This section is organized as follows:  first we present a comparison of some relevant
physical observables, namely, in Sec.~\ref{sec:Mass} the mass spectra, and in Sec.~\ref{sec:Decay} the decay constants. In
the same subsection \ref{sec:Decay}, an important consistency check is performed by computing decay constants
directly within the CST approach and comparing to those from the light-front formalism
using the CST mapped LFWFs. A discussion about the LFWFs themselves is given in
Sec.~\ref{sec:LFWFs}. We conclude with a calculation of several PDAs and PDFs in Sec.~\ref{sec:DA_PDFs} for
pseudoscalar and vector states.

 \subsection{Mass spectra}
  \label{sec:Mass}
The ability to reproduce the mass spectroscopy is a first test of meson models. It is known from the recent works in 
Refs.\cite{Li:2015zda,Leitao:2016bqq} that both approaches perform well in this regard. Here, we report the updated
spectra using the latest improvements of each model. A summary of the parameters used is given in Tables \ref{tab:BLFQmodel} and
\ref{tab:CSTmodel}. 

\begin{table}
\centering
\caption{BLFQ model parameters. $N_f$ is the number of flavors and $\mu_g$ is the gluon mass used to regularize the integrable Coulomb 
singularity.  $\kappa$ is the same as in Eqs.\ (\ref{eq:VLperp}-\ref{eq:VLx}) and $m$ is the constituent quark mass. $n_{\text{states}}$
is the number of states used in the fit. $N_{\max}=L_{\max}$.}
\begin{tabular*}{\columnwidth}{@{\extracolsep{\fill}}lllllll@{}}
\hline
\multicolumn{1}{@{}l}{} & $N_f$ & $\mu_g$ [GeV] & $\kappa$ [GeV] &$m$ [GeV] &$n_{\text{states}}$ & $N_{\max}$ \\
\hline
     $c\bar c$  & 4 & 0.02 & 0.966 & 1.603    &8 & \multirow{2}{*}{32} \\
    $b\bar{b}$  & 5 & 0.02 & 1.389 & 4.902   &14 &  \\
    \hline 
    $c\bar c$  & 4 & 0.02 & 0.985 & 1.570    &8 & \multirow{2}{*}{8} \\
    $b\bar{b}$  & 5 & 0.02 & 1.387 & 4.894   &14 &  \\
\hline
\end{tabular*}
\label{tab:BLFQmodel}
\end{table}

\begin{table}
\centering
\caption{CST model parameters with $N_{splines}=12$: $\Lambda$ is the Pauli-Villars cut-off parameter, used to regularize the UV behavior of both 
the Coulomb and linear potentials. $\sigma$ is the linear potential strength, $y$ is the mixing parameter defined in Eq.~(\ref{eq:kernel}), $\alpha_s$ is the
fixed quark-gluon coupling and $C$ is the Lorentz vector constant  of Eq.~(\ref{eq:V}). $m_q$ is the constituent quark mass and $n_{\text{states}}$ is the number of states
used in the fit (an extra $\bar{b}c$ state was also included in the fit).}
\begin{tabular*}{\columnwidth}{@{\extracolsep{\fill}}llllllll@{}}
\hline
\multicolumn{1}{@{}l}{} & $\Lambda$ & $\sigma$ [GeV$^2$] & $y$ & $\alpha_s$ &$C$ [GeV] &$m_q$ [GeV] & $n_{\text{states}}$ \\
\hline
$c\bar{c}$         & 2$m_q$ & 0.217 & 0.049 & 0.393 & 0.097 &  1.431&8 \\
$b\bar{b}$   & 2$m_q$  &0.217 & 0.049 & 0.393 &  0.097 & 4.786 &7 \\
\hline
\end{tabular*}
\label{tab:CSTmodel}
\end{table}

In the case of CST, the tensor mesons ($J\ge 2$) have not yet been calculated. Also as discussed in Ref.~\cite{Leitao:2016bqq}, the
axial-vector states do not have a definite $C$-parity. For that reason, they were not included in the fit. We observe a large relative deviation of $\chi_{c1}$ and $h_{c}$ when compared to other states. Both approaches predict consistently similar results for the yet unobserved states such as $3{}^1S_0$ and $1{}^3D_1$ and $2{}^3D_1$ in the bottomonium spectrum.

\begin{figure*}
 \begin{center}
    \includegraphics[width=.47\textwidth]{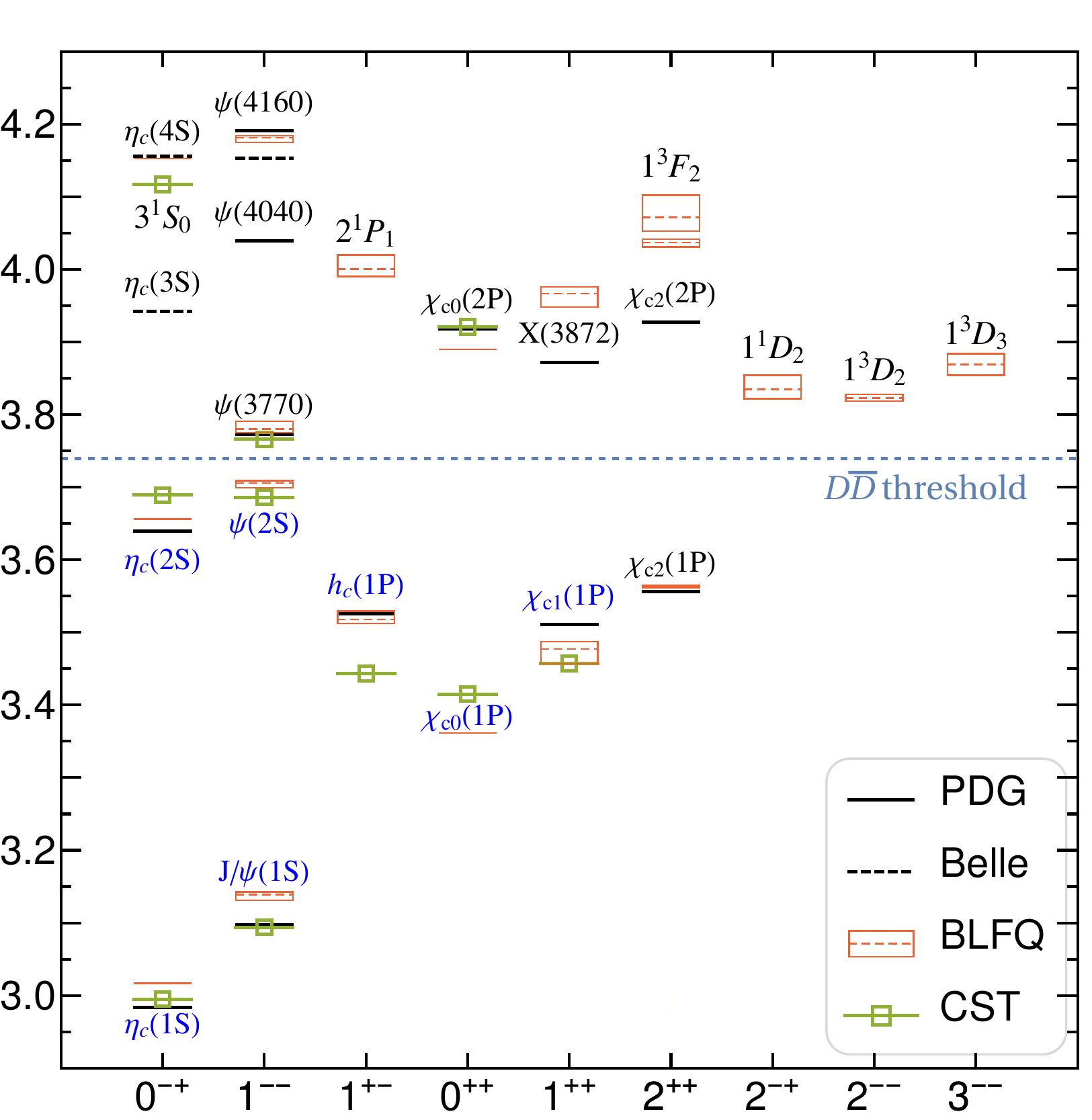} \qquad  
        \includegraphics[width=.48\textwidth]{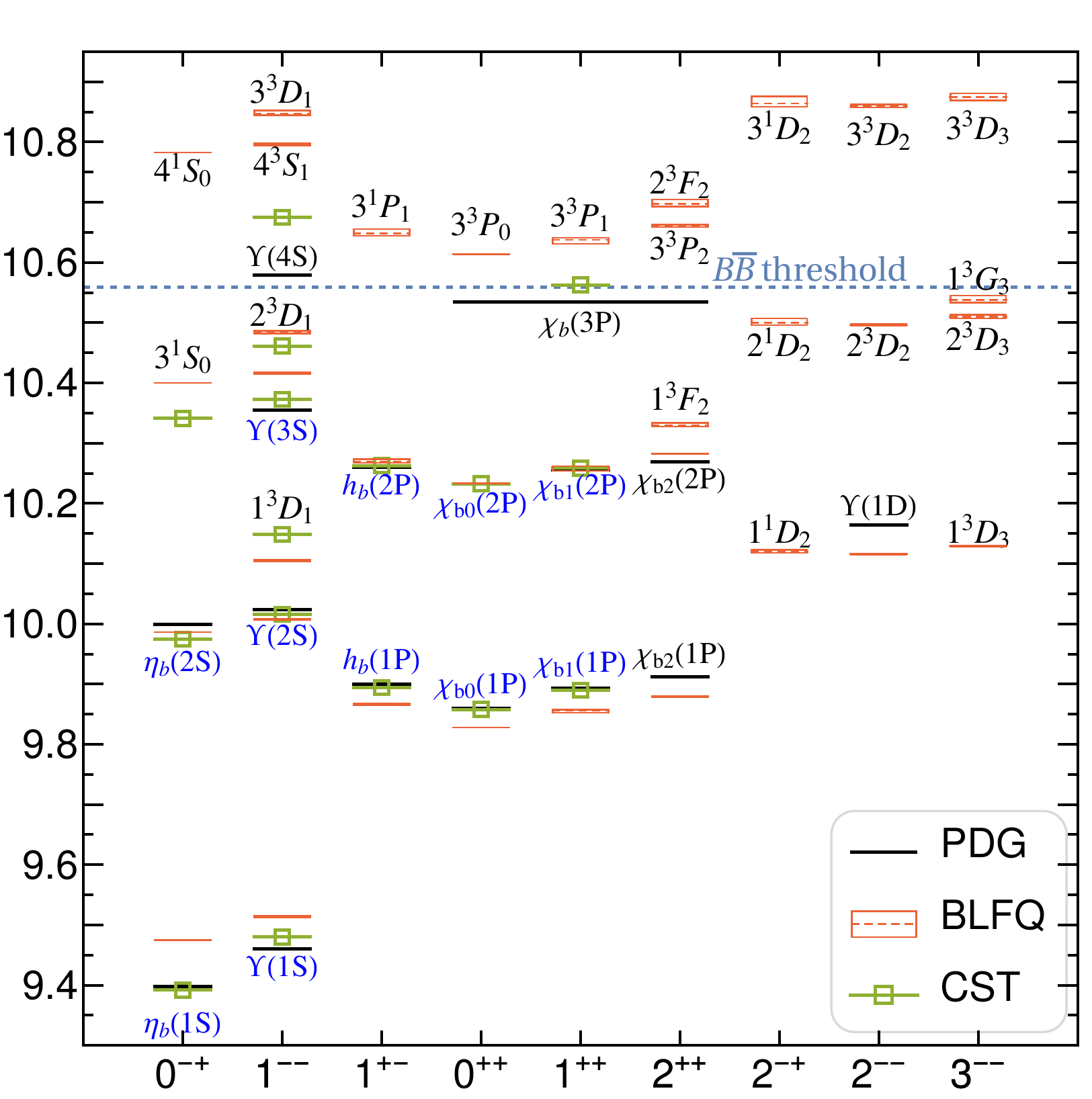}
  \caption{Charmonium (left) and bottomonium (right) mass spectrum from BLFQ and CST models compared with the experimental PDG values \cite{Agashe:2014kda}
 and BELLE  \cite{BELLE}  results. BLFQ states are marked by boxes to indicate the spreads of mass eigenvalues from different magnetic projections. The
mean values marked by dashed bars are defined as $\overline{M}\equiv [(M^2_{-J}+M^2_{1-J}+...+M^2_{+J})/(2J+1)]^{1/2}$, and $M_{\lambda}$ is the mass
eigenvalue associated with the magnetic projection $\lambda$, with $-J \le \lambda \le +J$.}\label{fig:massBBCC}
\end{center}
\end{figure*}

 Finally, in order to quantify the agreement of the predictions with the experimental measurements, we
determined the root-mean-square difference to the measured states below threshold and shown in
blue in Fig.\,\ref{fig:massBBCC}. The results are $\delta_{rms}$ (BLFQ)=39 MeV  and $\delta_{rms}$(CST)=11 MeV for 11 bottomonium states. The results for charmonium are $\delta_{rms}$ (BLFQ)=33 MeV and $\delta_{rms}$(CST)=42 MeV with 7 states.
 
\subsection{Decay constants}
\label{sec:Decay}

\begin{figure*}
  \centering
    \includegraphics[width=.5\textwidth]{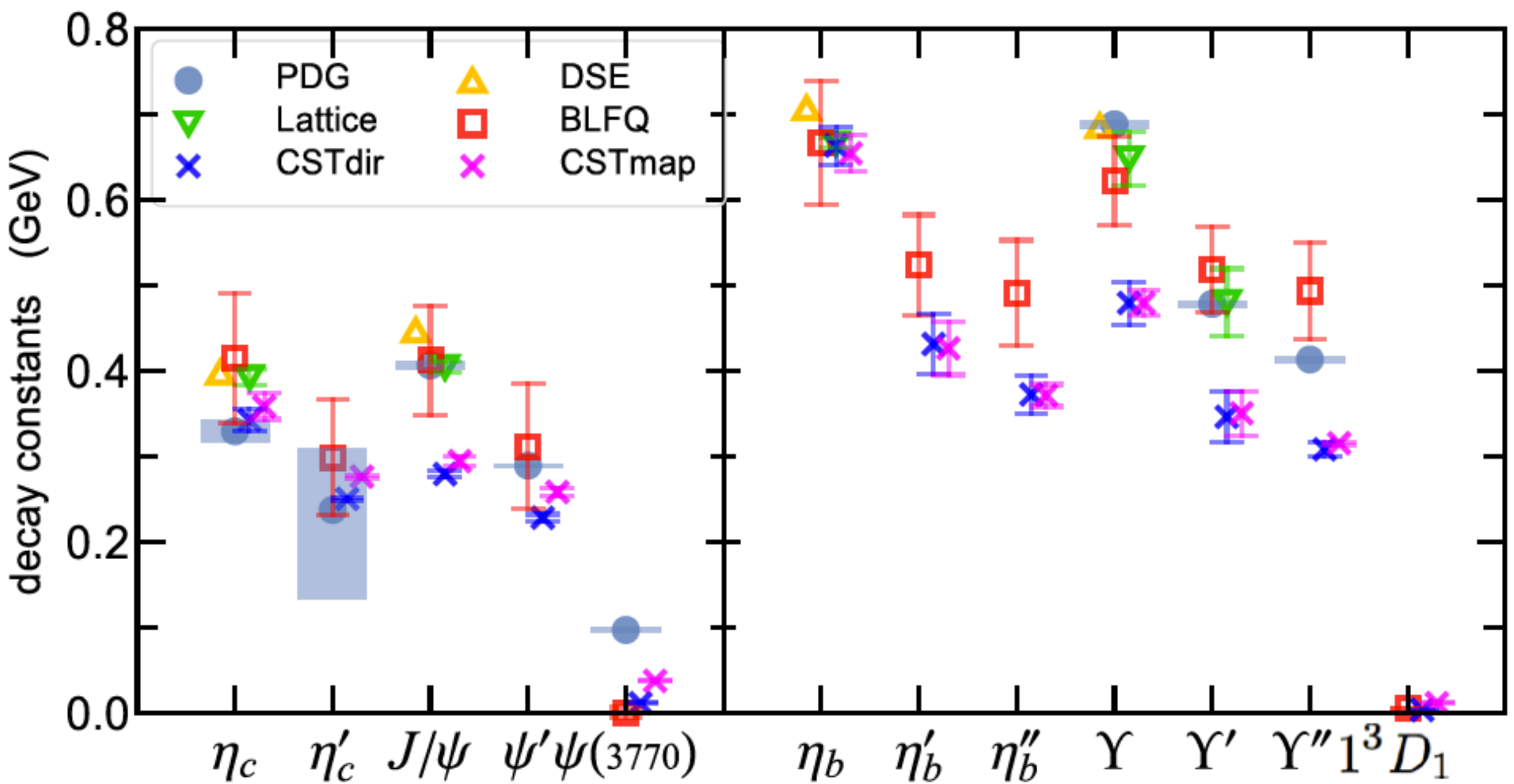}
  \caption{Comparison between the decay constants determined with CST-LFWFs and with BLFQ-LFWFs as well as with other methods and with experiment (PDG). The lattice results are from Ref.~\cite{Davies:2010ip,McNeile:2012qf,Donald:2012ga,Colquhoun:2014ica} and the DSE results are from Ref.~\cite{Blank}.}
  \label{fig:compdecays}
\end{figure*}

Using the previous definitions we calculate the bottomonia and charmonia decay constants for pseudoscalar and vector states. The
results are shown in Fig.\,\ref{fig:compdecays} and compared with experimental PDG values \cite{Agashe:2014kda}, as well as
lattice QCD \cite{Davies:2010ip,McNeile:2012qf,Donald:2012ga,Colquhoun:2014ica} and
Dyson-Schwinger equations (DSE) \cite{Blank} predictions. 

In BLFQ the results were obtained with $N_{\text{max}}=8$ for charmonium and $N_{\text{max}}=32$ for bottomonium, in order to guarantee the aforementioned UV cut-off of $1.7m_q$. The theoretical uncertainty is
estimated by varying the scales: for charmonium $\Delta f = 2|f_{{\max}=8} - f_{{\max}=16}|$; for bottomonium $\Delta f = 4|f_{{\max}=24} -f_{{\max}=32}|$. As estimates for the numerical uncertainties of the CST calculations we took the difference between the results obtained with 12 and 8 splines.

Overall, both CST and BLFQ results reproduce the data quite well and are consistent with other approaches, where available.  In the CST approach the pseudoscalar decay constants are closer to experiment than the vector meson decay constants. It is worth emphasizing that these results are pure predictions in the sense that none of them were included in the fits. For that reason the agreement with experimental data for decay constants (where available) could be improved by incorporating those data in the model parameter fits. The only notable discrepancy comes from the $D$-wave {\emph{vector} mesons} $\psi(3770)$ and ${1}^3D_1$ where the decay constants are very small in both approaches, as is also observed from some other
approaches but not all (see Ref.~\cite{Krassnigg:2016hml} and the references therein). In principle, these decay constants
are only non-vanishing due to the mixing with the $S$-wave in the non-relativistic case. However, $\psi(3770)$ is just above
the open-charm threshold. It will be interesting to resolve the theoretical speculations in $1{}^3\!D_1$ from experimental measurements.
\\

Next we discuss an important test of the map adopted and described in section \ref{sec:BHLmap}. The CST solutions as functions of $\bf k$ as in Eq.~(\ref{eq:CSTamplitudes}) are first expressed in terms of $x$ and $k_\perp$ using Eq.~(\ref{eq:BHLx}), and then we use the light-front definitions of Eqs.~(\ref{eq:BLFQdecayp}--\ref{eq:BLFQdecayv}) to
recalculate the decay constants.

\begin{table}
\centering
\caption{Decay constants with CST amplitudes as LFWFs (map.) and original CST amplitudes (dir.) and absolute difference $\delta$ in units of MeV.}
\begin{tabular*}{\columnwidth}{@{\extracolsep{\fill}}llll | llll@{}}
\hline
\multicolumn{1}{@{}l}{}   &map. & dir. & $\delta$ && map. & dir. & $\delta$ \\
    \hline
  $\eta_c$ & 359(10) & 343(9)  &16 &  $\eta_b$ & 655(14) & 664(15)  &9\\
  $\eta'_c$& 277(2) & 251(2)  &26  &    $\eta'_b$& 427(21) & 432(23)  &5\\
       & &  &  &   $\eta''_b$ & 372(9) & 373(15)  &1 \\
   $J/\psi$& 295(4) & 280(3)  &15 & $\Upsilon$& 480 (10)& 480(17)  &0\\
    $\psi'$& 259(3) & 229(3)  &30& $\Upsilon'$& 351(18) & 347(20)  &4  \\
     & &  &  &    $\Upsilon''$& 316(2) & 309(6)  &7 \\
    $\psi(3770)$& 38(1) & 12(1)  &26 &  $1 ^{3}D_1$& 12(1) & 4(1)  &8 \\
\hline
\end{tabular*}
\label{tab:decayCSTmap}
\end{table}

These new ``CST-mapped'' results (indicated by ``CSTmap'') are presented in Fig.\,\ref{fig:compdecays} as well. For bottomonium states, shown in the right panel of Fig.\,\ref{fig:compdecays}, the difference between the two sets of calculations is smaller than the numerical errors.
For charmonium, on the left panel, the decay constants determined with the CST-LFWFs are slightly larger. The absolute differences $\delta$ are
listed in Table \ref{tab:decayCSTmap}. The largest deviation (disregarding the $D$-wave vector states which have tiny decay
constants) does not surpass 30 MeV (about 10\% in charmonium) and 9 MeV (2\% in bottomonium), confirming that the BHL prescription 
we use works better as one approaches the non-relativistic limit. Nevertheless, this test provides a reasonable justification for the procedure we follow in CST to obtain heavy 
quarkonia LFWFs, which we will review in more detail in the next section.

\subsection{Light-front wave functions}
\label{sec:LFWFs} 

Having new sets of light-front wave functions for quarkonia derived from the CST approach opens the door for several calculations. As already mentioned, the CST equation solved in this work does not respect charge-conjugation symmetry, and thus the CST wave functions do not have a definite $C$ parity. A direct comparison with the BLFQ solutions, which do have definite $C$ parity, allows for a better identification of the axial-vector states obtained from CST. From the BLFQ side, a direct comparison with LFWFs from a different approach also offers benefits. As mentioned earlier, in BLFQ the inevitable basis truncation breaks the rotational symmetry. The total angular momentum $J$ is not well defined and the state identification is based on
spectroscopy with the help of $P$, $C$, etc. Comparing with CST results, for which $J$ is an exact quantum number, gives guidance to validate this identification. With their rich radial and angular structure, the bottomonium vector meson LFWFs for
instance provide a non-trivial test of the methods for identifying $J$ in the BLFQ results.
 
We investigated LFWFs of all states below open flavor thresholds and with $J<2$ (cf. Fig.\, \ref{fig:massBBCC}) and for  all
non-vanishing spin configurations. The obtained wave functions exhibit close
correspondence between CST and BLFQ in their dominant structures for all states and spin alignments. 

To visualize the rich structures of the wave functions, we adopt the scheme of Ref.~{\cite{LiLC2016}}. We note that for a particular polarization $\lambda$ and spin alignment $s\bar{s}$, the LFWFs can be expressed as
\begin{equation}
\psi_{s\bar{s}}\left({\bm k}_\perp,x\right) = \Phi_{s\bar s} \left({k}_\perp,x\right)\exp(i m_\ell \phi),
\end{equation} 
 where $k_\perp = |\bm{k}_\perp|$ and $\phi = \arg \bm{k}_\perp$. This is valid because the orbital angular momentum projection
$m_{\ell}=\lambda-s-\bar{s}$ is definite ($\lambda \equiv m_J$). In order to visualize these wave functions, we drop the phase $\exp(i
m_{\ell}\phi)$, while retaining the relative sign $\exp(i m_\ell\pi)=(-1)^{m_\ell}$ for negative values of $k_\perp$. More precisely we plot
\begin{equation}
\Psi \left({k}_\perp,x\right)\equiv \left\{
                \begin{array}{lr}
                  \Psi \left({k}_\perp,x\right), & k_\perp  \geq 0,\\
                 \Psi \left(-{k}_\perp,x\right)(-1)^{m_\ell}, & k_\perp < 0.
                \end{array}
              \right.
\end{equation}
This scheme essentially takes a slice of the 3D wave function $\psi_{s\bar s}(\bm k_\perp, x)$ at $k_y = 0$.

\begin{figure}
 \begin{center}
    \includegraphics[width=.45\textwidth]{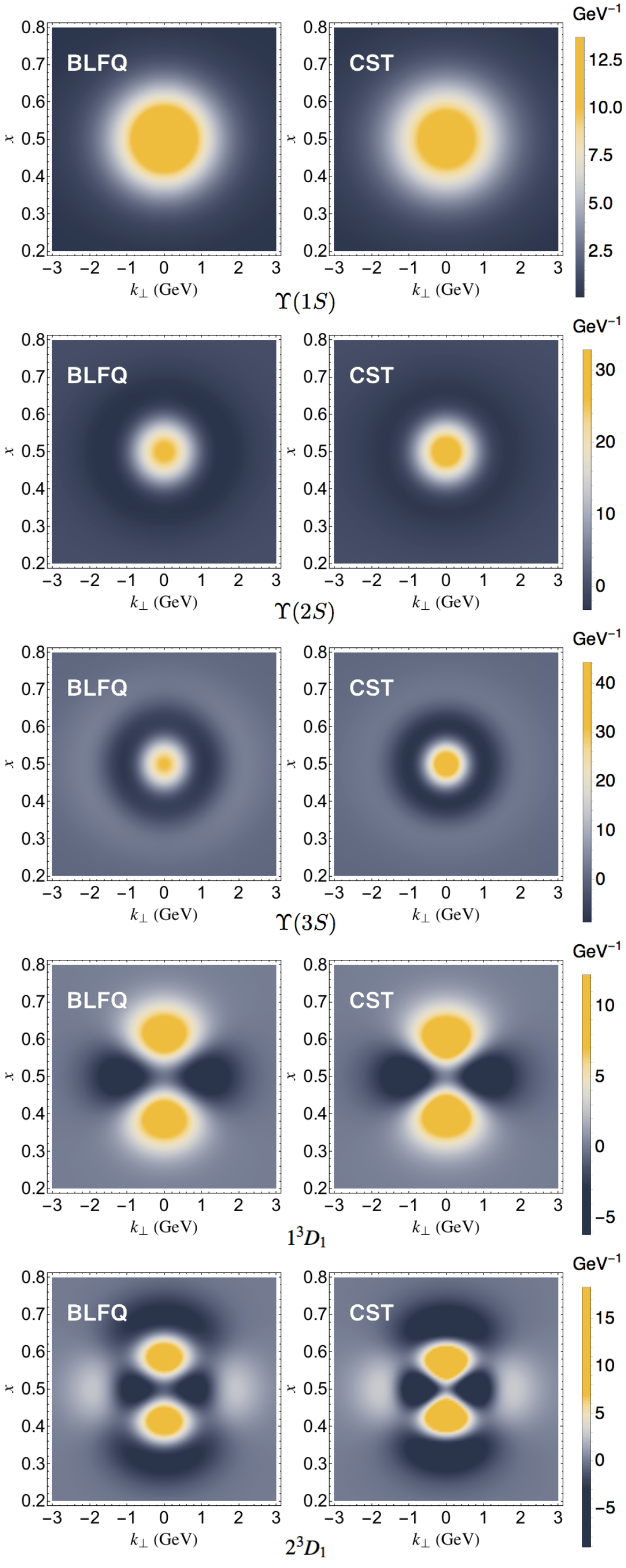}
  \caption{Dominant triplet component of the BLFQ-LFWFs and CST-LFWFs for several bottomonium vector meson states. Both plots have the same scale and the region outside $x=0.2$ and $x=0.8$ is not depicted because it is structureless.}
\label{fig:upsi_states}
\end{center}
\end{figure}

Let us begin the discussion of the LFWFs with the interesting case of the vector $b\bar{b}$ because from all the systems this is the one
with the largest number of states below its open flavor threshold, the $B \overline B$ threshold. These systems admit a mixture
of $S$- and $D$-wave components (as long as there is a tensor force). In Fig.\,\ref{fig:upsi_states} we show the dominant triplet
component of the ground state and several radial excitations. The states $\Upsilon(1S)$, $\Upsilon(2S)$, and $\Upsilon(3S)$ are clearly
$S$-wave dominated and in both cases an increasing number of nodes in both transverse $(k_\perp)$ and longitudinal ($x$)
directions is observed. As a consequence, and for our particular choice of the coordinate range, a nesting ring
pattern emerges. 
This is consistent with the non-relativistic interpretation, where the radial excitation is homogeneous in all three directions. Prior to the
map described in Eq.~(\ref{eq:BHLx}), CST amplitudes expressed as functions of ${\bm k}$ show precisely this behavior (see Fig.\ 3 of
Ref.~\cite{SLquarkconf}). The $1{}^3\!D_1$ wave function resembles the shape of the the spherical harmonic $Y^0_2(\hat{\bm k})$. The
same happens for $2{}^3\!D_1$, where the complicated inner structure is also compatible with a $Y^0_2(\hat{\bm k})$ but now with an extra node
in both $k_\perp$ and $x$.

\begin{figure}
\captionsetup[subfloat]{singlelinecheck=off}
\centering
  \subfloat[]{%
    \includegraphics[width=.47\textwidth]{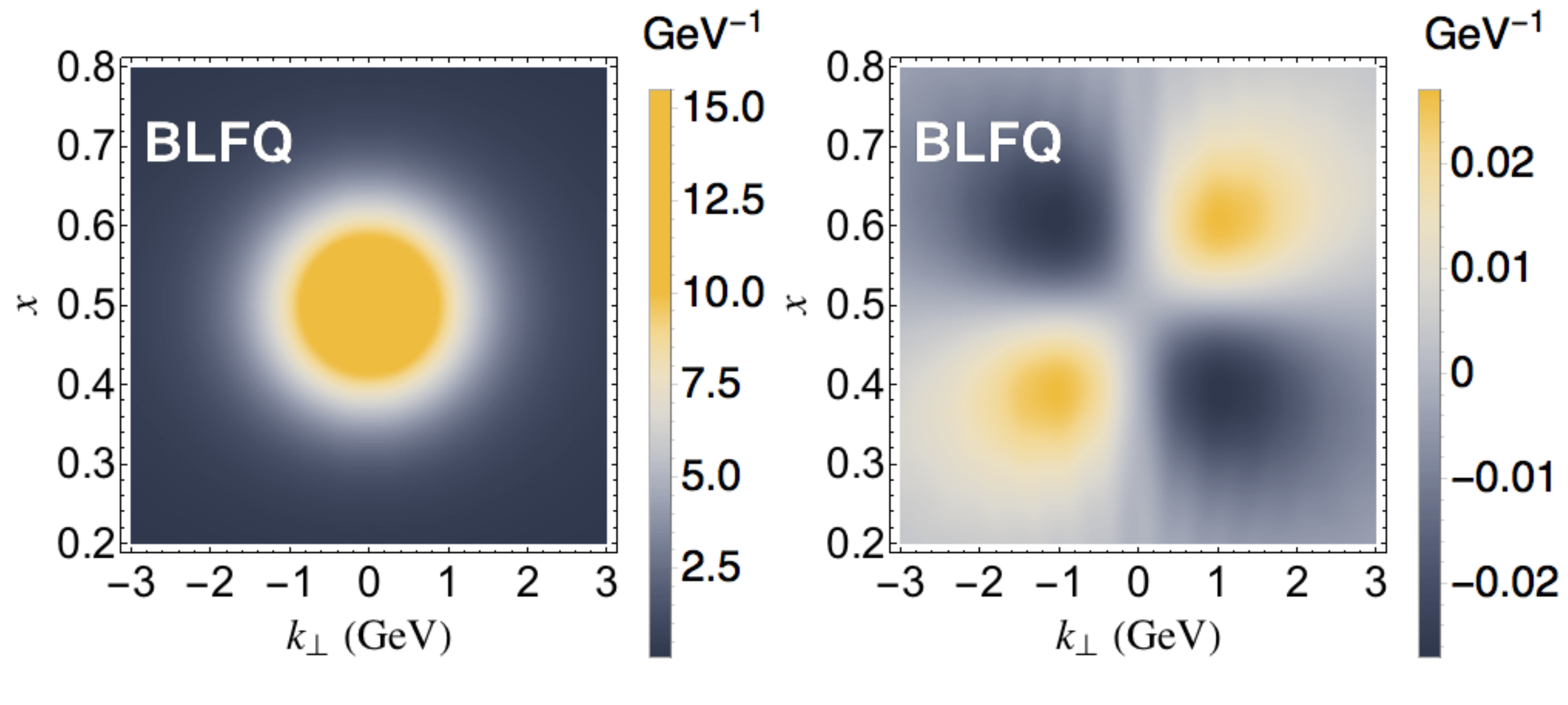}}\\
  \subfloat[]{ \includegraphics[width=.48\textwidth]{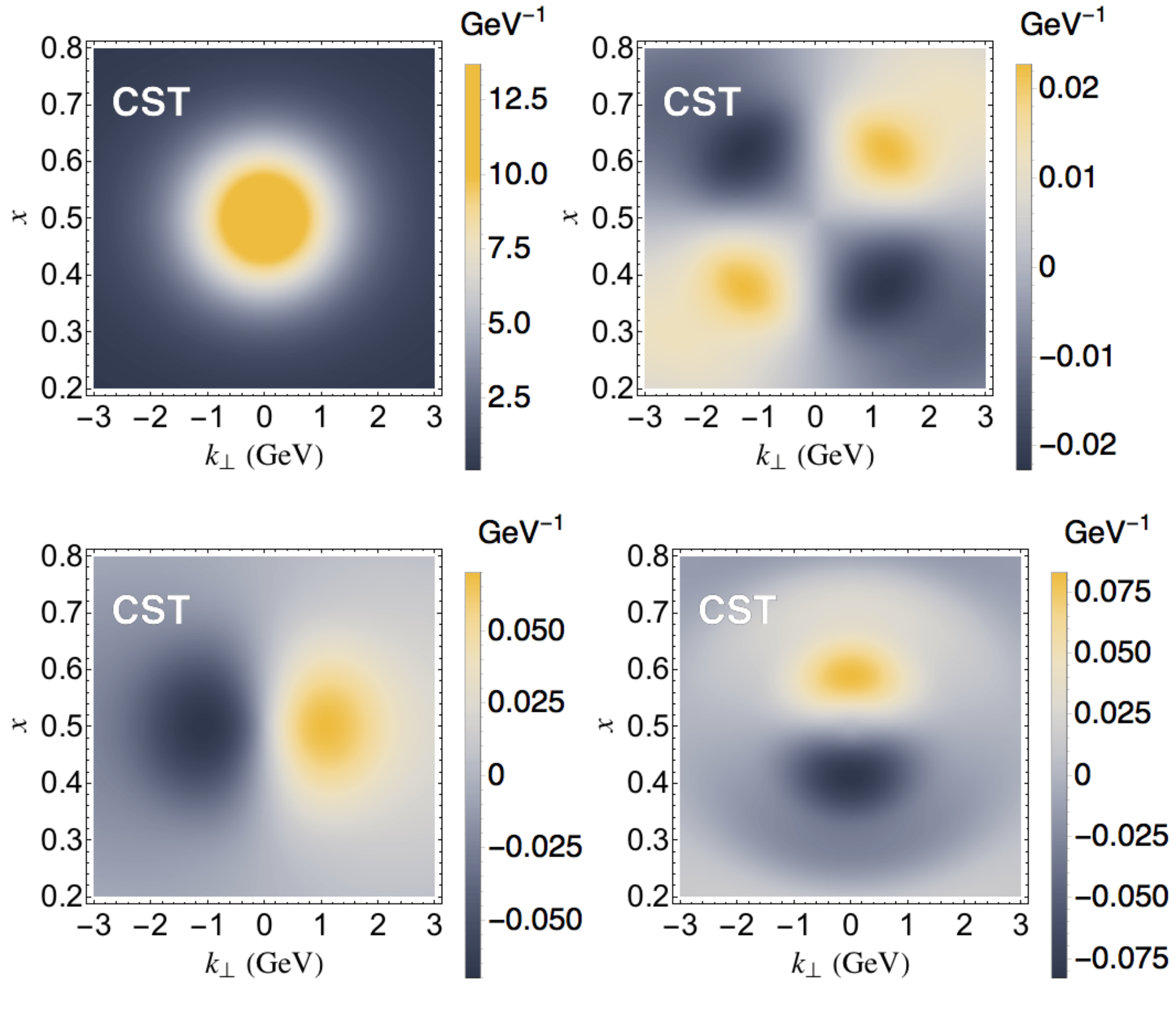}}
  \caption{Complete set of light-front wave functions of $\Upsilon(1S)$ with $\lambda=0$. (a) BLFQ results:  dominant triplet component  $\psi_{(\uparrow\downarrow+\downarrow\uparrow)}$ (left)  and 
subdominant $\psi_{\downarrow\downarrow}$ (right); (b) CST results: dominant triplet component $\psi_{(\uparrow\downarrow+\downarrow\uparrow)}$  of the LFWF (top left) and 
$\psi_{\downarrow\downarrow}$ (top right), both from $\psi^{+-}$. Subdominant
components $\psi_{\downarrow\downarrow}$ (bottom left) and singlet $\psi_{(\uparrow\downarrow-\downarrow\uparrow)}$ (bottom right), both from
$\psi^{++}$.\label{fig:upsi_all_components_b}}
  \label{fig:upsi_all_components}
\end{figure}

In Fig.\,\ref{fig:upsi_all_components}, in addition to the dominant triplet component, other sub-dominant components of purely relativistic origin are shown. Here, significant differences appear between
the CST and the BLFQ LFWFs. While in BLFQ there is only one $\psi_{\downarrow\downarrow}$ component, in CST two extra components compatible
with a quantum number $\ell=1$ and with spin alignments $\psi_{\downarrow\downarrow}$ and spin singlet
$\psi_{(\uparrow\downarrow-\downarrow\uparrow)}$ appear and are presented in the last row of Fig.\,\ref{fig:upsi_all_components_b}. These components emerge from the CST amplitude's $\psi^{++}$ component [cf. Eq.~(\ref{eq:CSTamplitudes})] and are absent in BLFQ
because there positive energy and negative energy states do not mix.

\begin{figure*}
\centering
\includegraphics[width=.9\textwidth]{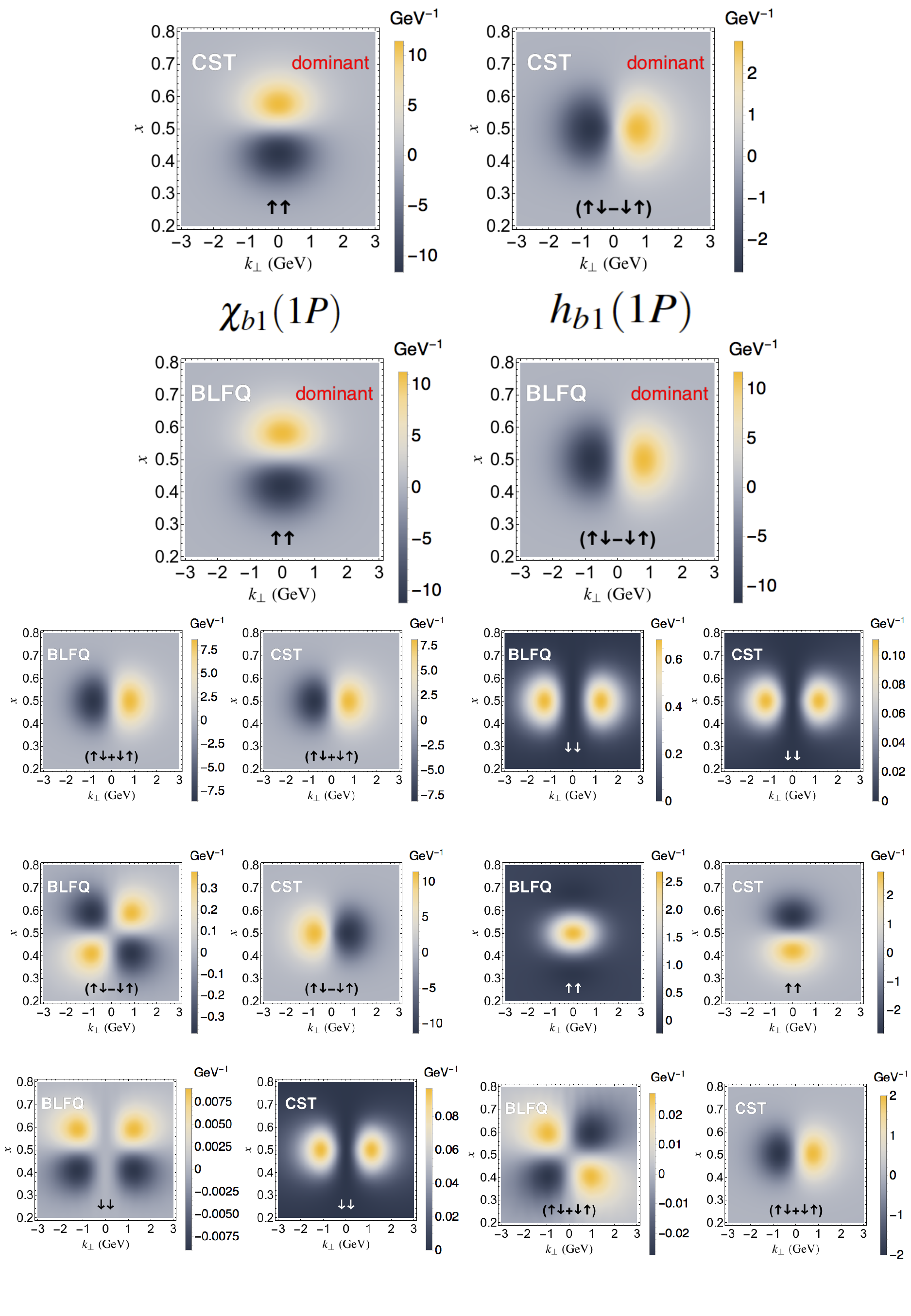}
\caption{LFWFs for the two lowest mass axial-vector states of bottomonium with polarization $\lambda=1$. The dominant of the mapped CST LFWFs, labeled ``dominant'', are identified based on the matching dominant BLFQ LFWFs component, $\chi_{b1}$(1P) on the left and $h_{b1}$(1P) on the right. In the lower half of the figure we show subdominant LFWFs components of BLFQ and CST side by side,  $\chi_{b1}$(1P) on the left and $h_{b1}$(1P) on the right. The arrows inside each individual panel represent a given spin-alignment.}
  \label{fig:Cparityfig}
\end{figure*}

In order to demonstrate how the comparison of the mapped CST with the BLFQ LWFWs can help in the identification of states we give an example: in Fig.\,\ref{fig:Cparityfig} we show the dominant components (always labeled ``dominant'') of the two lowest mass solutions of the bottomonium $1^+$ states in both formalisms. We observe that the mapped spin triplet CST LFWF resembles very closely the dominant triplet BLFQ component $\psi^{\lambda=1}_{\uparrow\uparrow}$ (on the left) with $J^{PC}=1^{++}$, whereas the spin singlet CST LFWF has its correspondence in the dominant singlet BLFQ component $\psi^{\lambda=1}_{(\uparrow\downarrow-\downarrow\uparrow)}$ (on the right) with $J^{PC}=1^{+-}$. Therefore we can conclude that these two CST LFWFs describe mostly the $\chi_{b1}(1P)$  and the $h_{b1}(1P)$ states, respectively. However, a comparison of the sub-dominant components shows differences: for instance, one of the BLFQ components of the $\chi_{b1}$(1P) state, $\psi^{\lambda=1}_{\downarrow \downarrow}$, exhibits $F$-wave features, which, in principle, violates the angular momenta addition $|L-S| \le J$. Such $F$-wave contributions are 
 not present in the CST LFWFs for states with $J=1$. On the other hand, the CST LFWF with the same spin alignment, $\psi^{\lambda=1}_{\downarrow \downarrow}$, is a $D$-wave, which in BLFQ solutions only appears for the $h_{b1}(1P)$ state. Nevertheless, in the near future, we will solve the more complicated CST equations with charge conjugation symmetry and  some of the observed differences are expected to disappear.  

 \subsection{Leading-twist parton distribution amplitudes and parton distribution functions}
 \label{sec:DA_PDFs}
In this section we present the results for the leading-twist parton distribution amplitudes of $^1 S_0$ and $^3 S_1$ states and compare them with other results from the literature. For completeness and in order to compare with other approaches, the moments are displayed in Table \ref{tab:moments_DA}. In addition, estimates for the root mean square velocity of the constituents are displayed in Table~\ref{tab:rmsvelocity}. 

  \begin{figure*}
 \begin{center}
    \includegraphics[width=.35\textwidth]{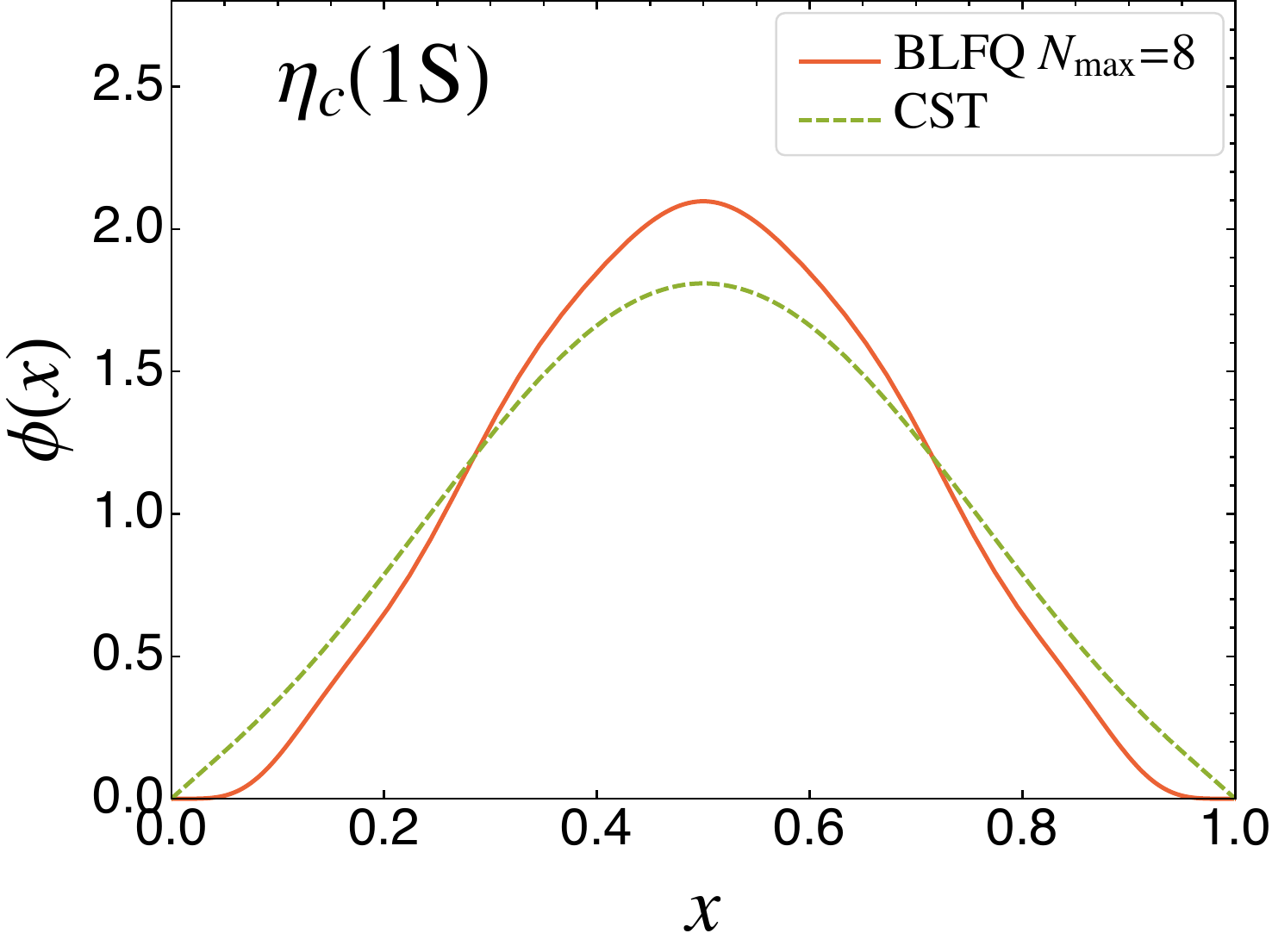}
  \includegraphics[width=.35\textwidth]{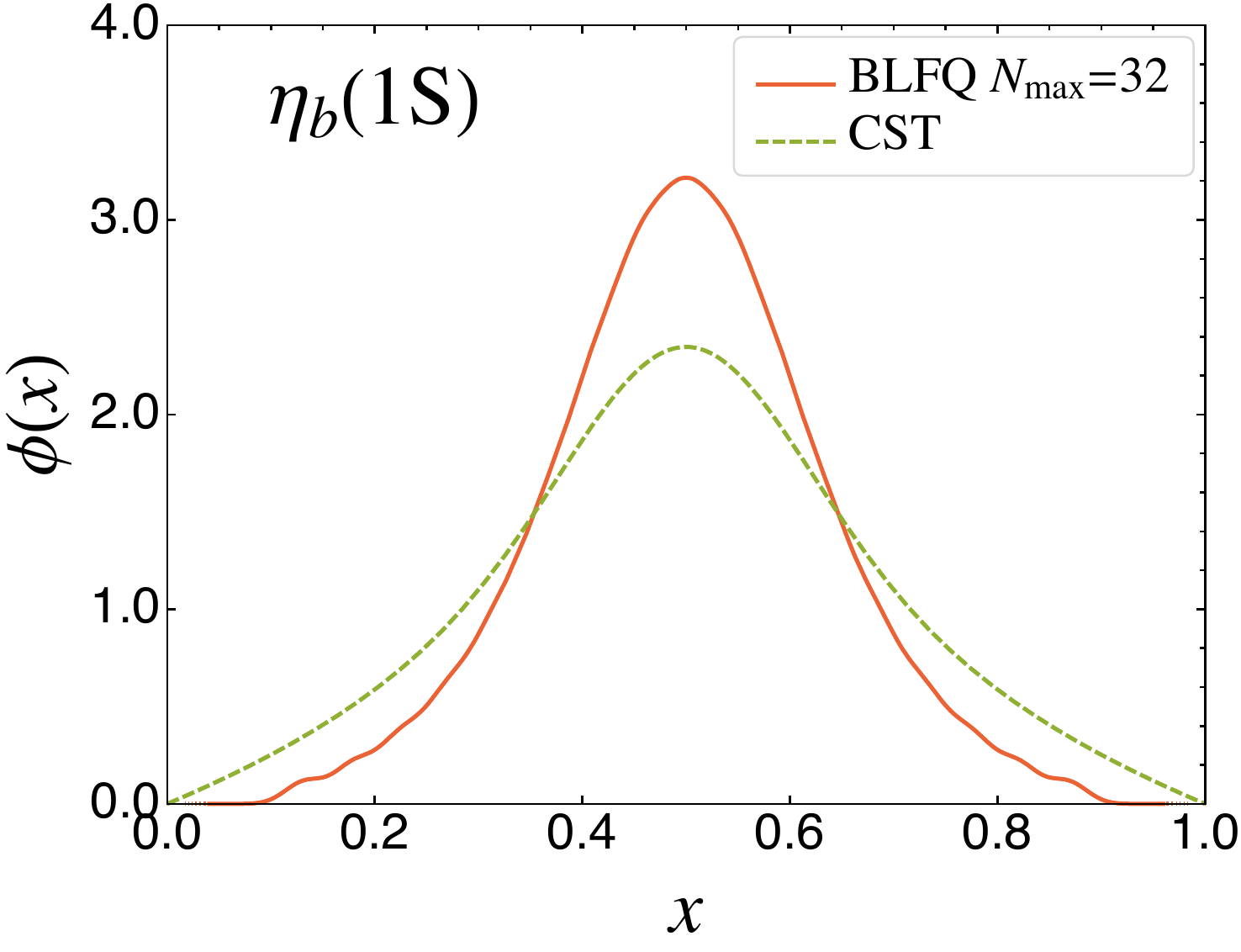}\\
      \includegraphics[width=.35\textwidth]{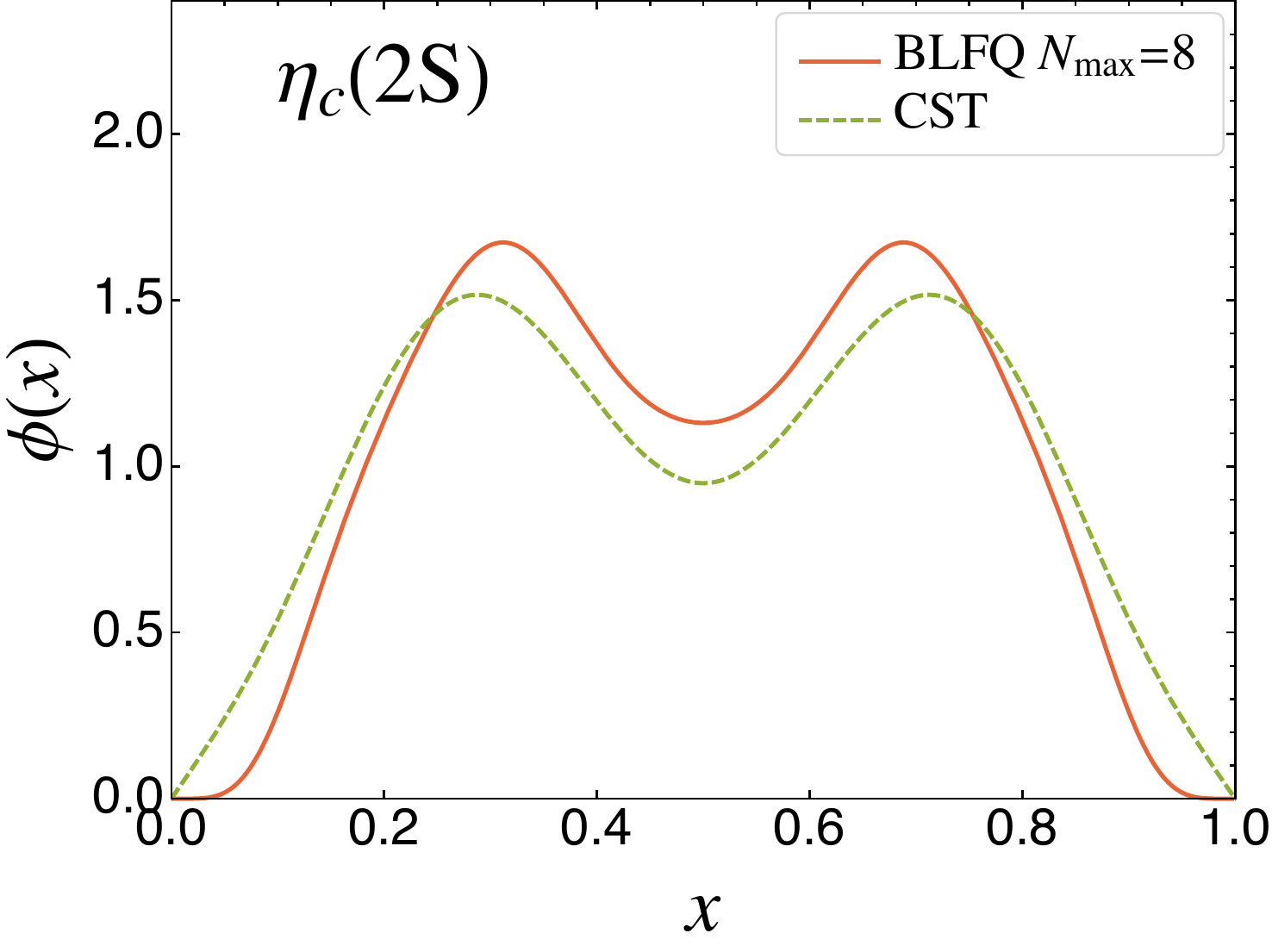}
  \includegraphics[width=.35\textwidth]{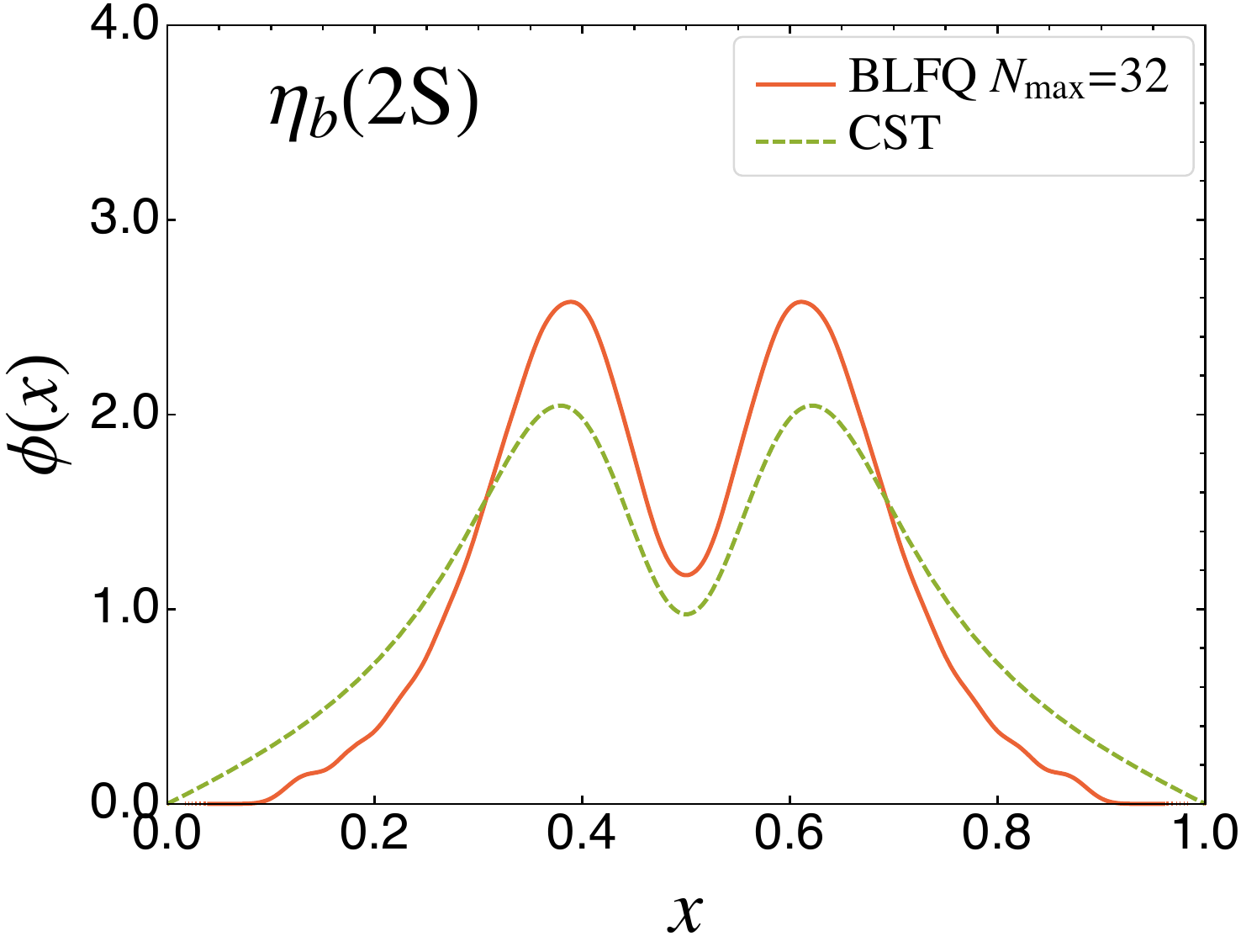}\\
      \includegraphics[width=.35\textwidth]{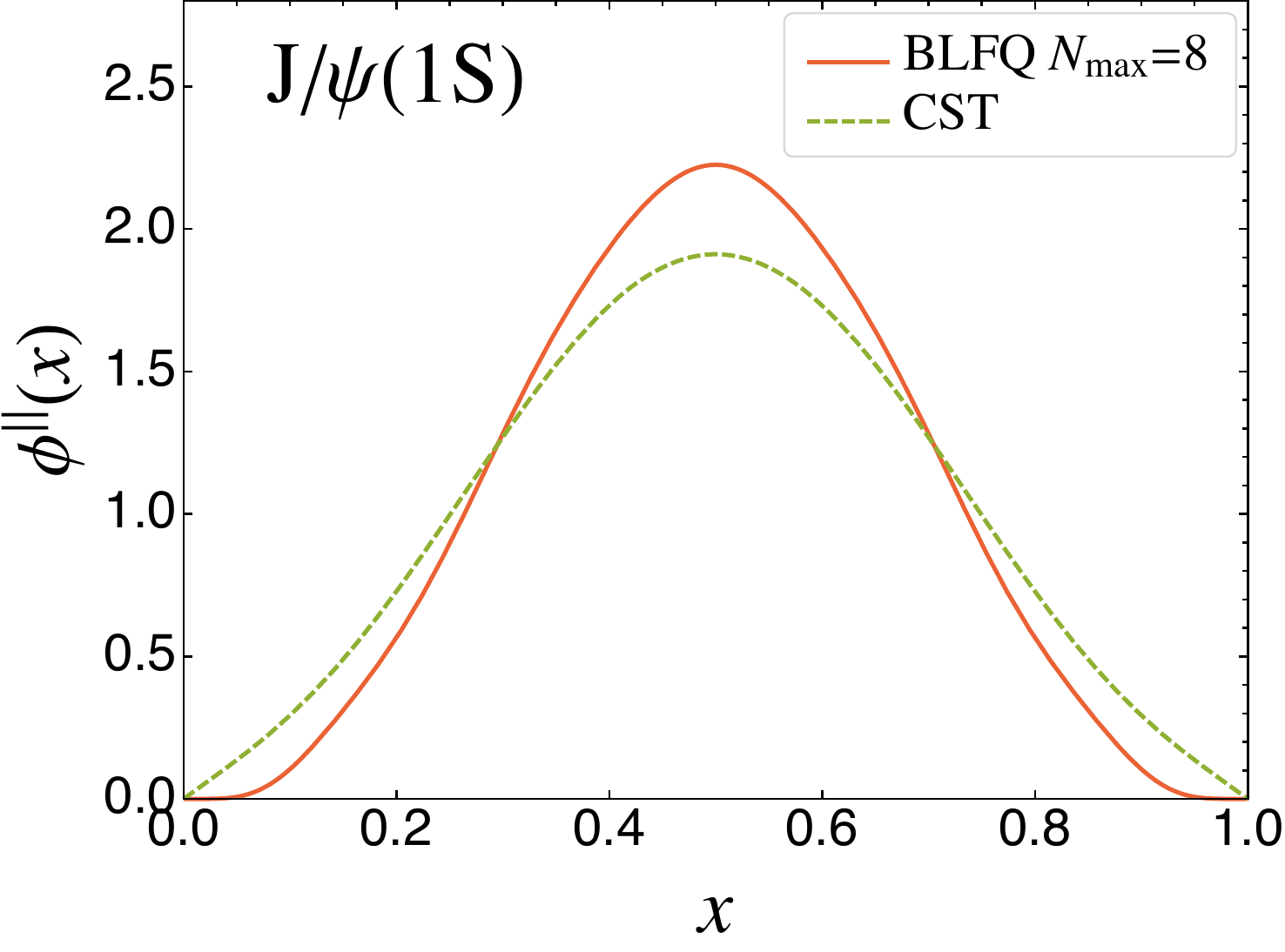}
      \includegraphics[width=.35\textwidth]{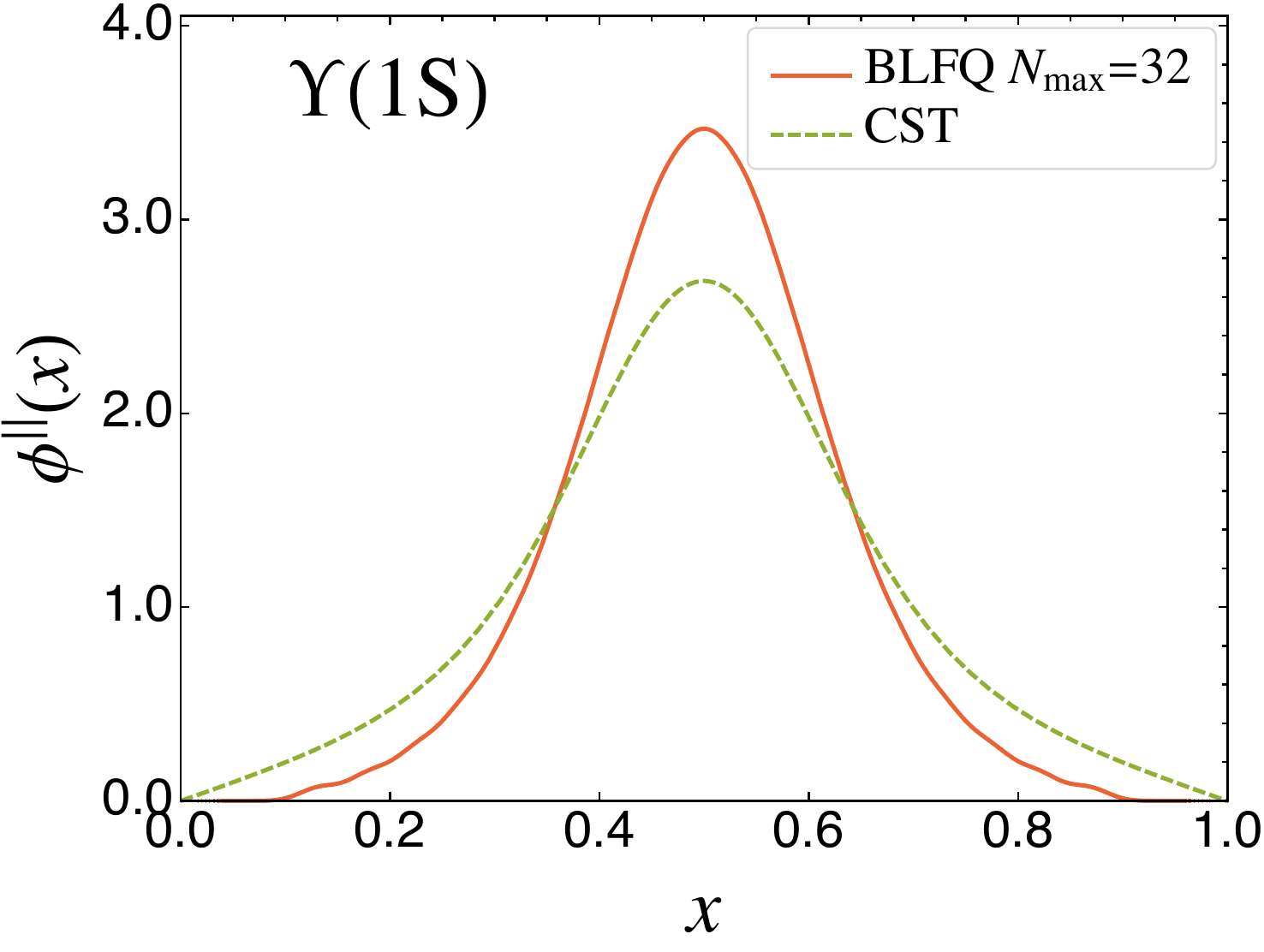}\\
      \includegraphics[width=.35\textwidth]{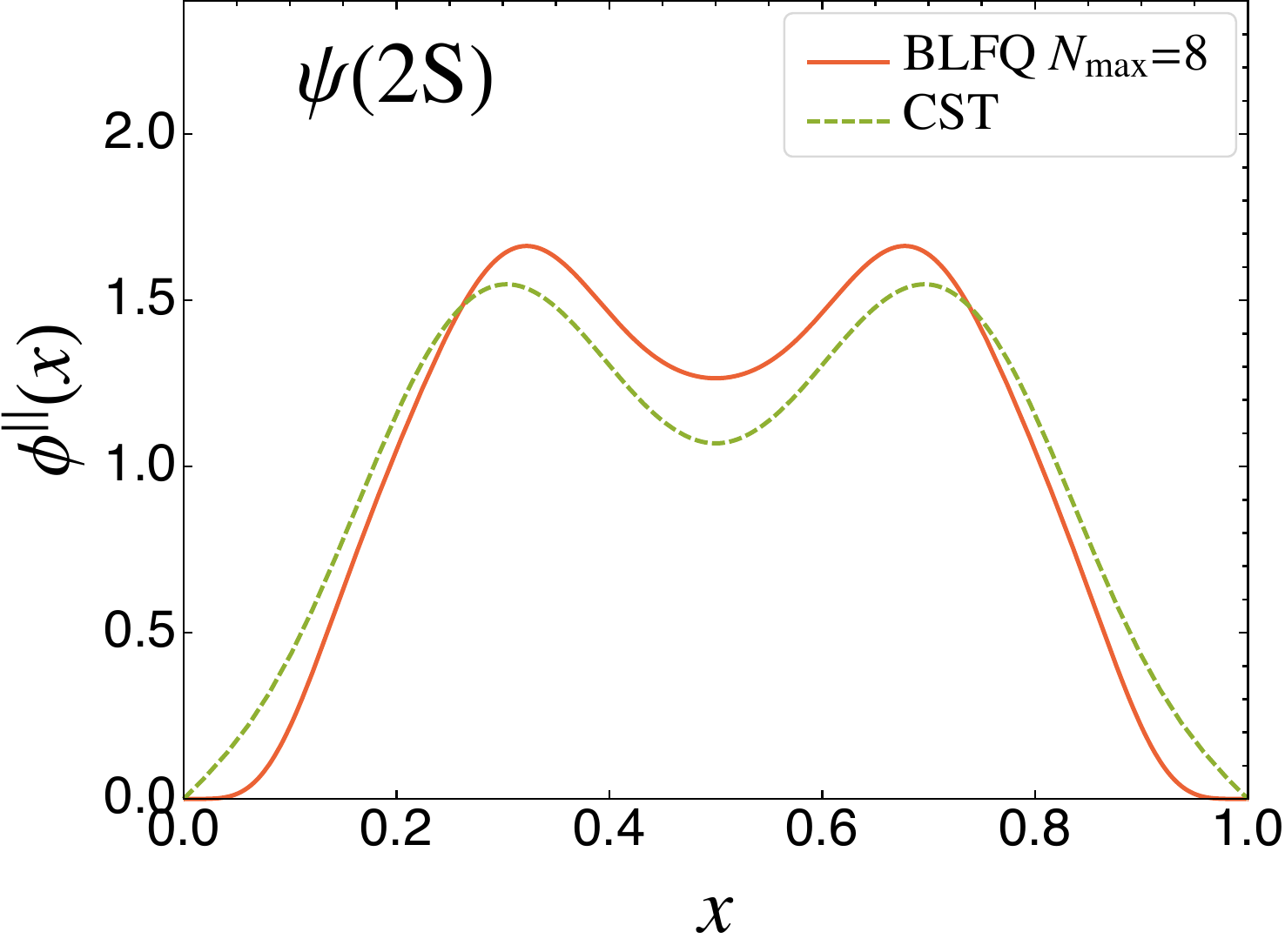}
      \includegraphics[width=.35\textwidth]{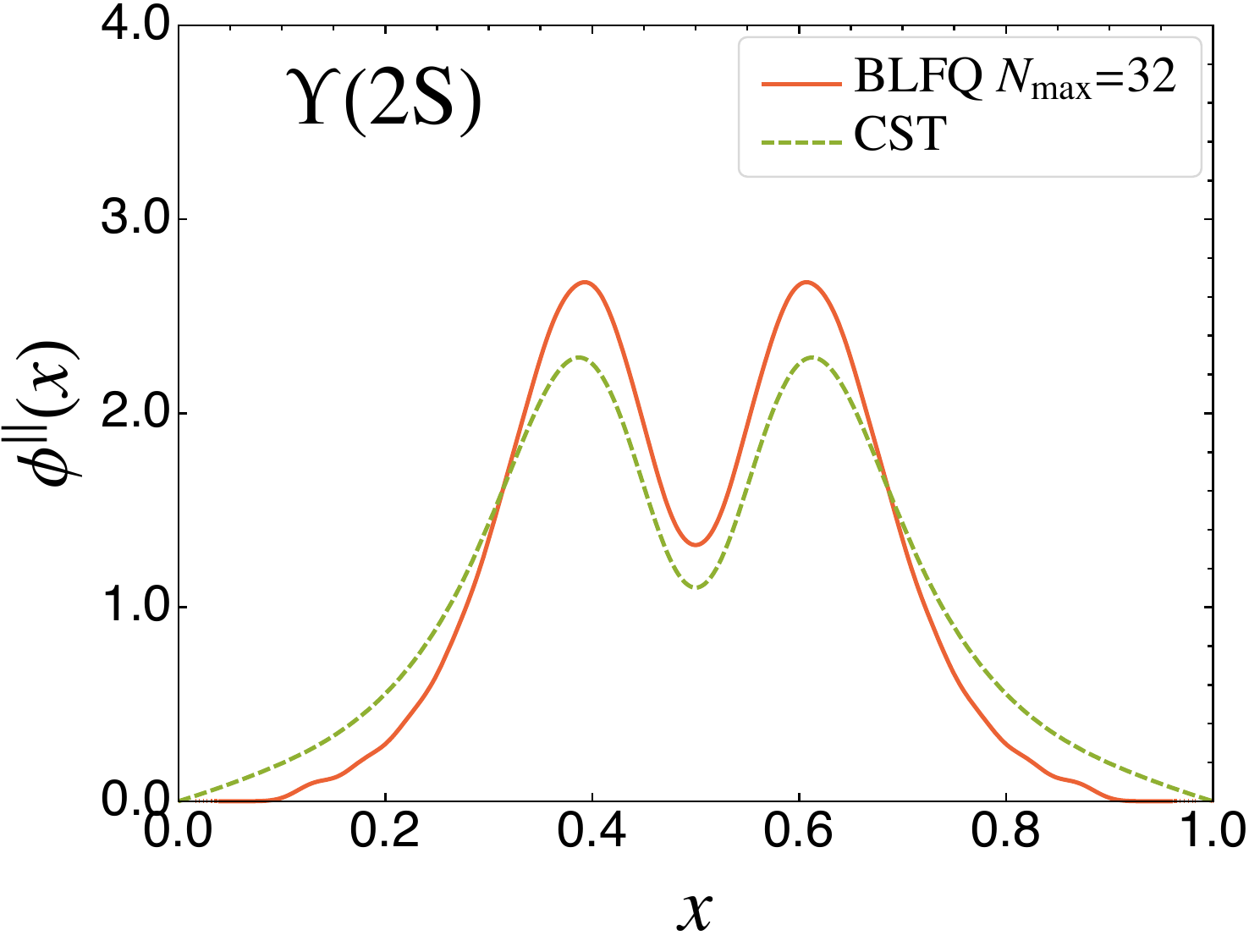}\\
  \caption{Distribution amplitudes for pesudoscalar and vector states.}
  \label{fig:DAs}
\end{center}
\end{figure*}
In Fig.\,\ref{fig:DAs}, distribution amplitudes are given for pseudoscalar and vector ground states and their corresponding first radial excitations. 

Several global features can be observed in both approaches, such as the number of maxima and minima and the consistent broader curves for charmonium than for bottomonium.  We notice however that the CST distribution amplitudes exhibit larger tails than the BLFQ distribution amplitudes where a gaussian behavior causes a stronger falloff. This indicates that in the CST approach the constituent quarks are noticeably more relativistic as suggested by the estimate of $\langle v^2 \rangle$ (cf. Table \ref{tab:rmsvelocity}).

\begin{figure}
 \begin{center}
    \includegraphics[width=.39\textwidth]{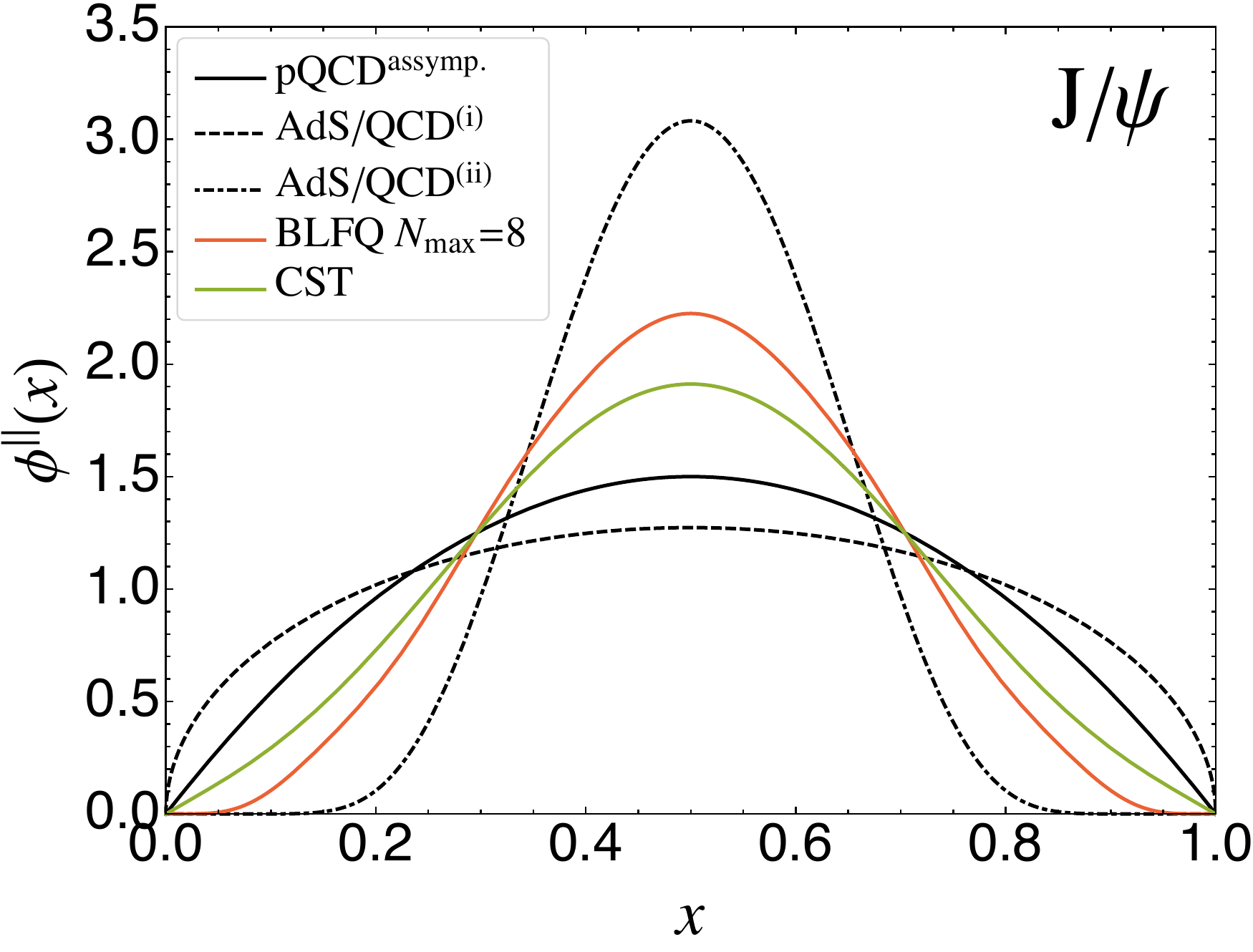}
  \caption{Comparison of CST and BLFQ longitudinal leading twist distribution amplitude of $J/\psi$ with respect to pQCD${}^\text{asymp}$ prediction \cite{Lepage1980}; AdS/QCD${}^\text{(i)}$ of Brodsky and de T\'eramond \cite{Brodsky:2014yha} and AdS/QCD${}^\text{(ii)}$ of Swarnkar {\it et al.} \cite{Swarnkar}.}
\label{fig:DA_jpsi}
\end{center}
\end{figure}
 
\begin{table*}
\caption{Moments $\langle  \xi^n \rangle$ of leading-twist PDAs for pseudoscalar and vector states with longitudinal polarization. The BLFQ results are given at $N_{\text{max}}=8$ for charmonium and $N_{\text{max}}=32$ for bottomonium. The other moments correspond to the results from NRQCD \cite{Bodwin}, QCD sum rules  \cite{BragutaPLB} and DSE \cite{DSE2016}. The pQCD asymptotic value for $\langle  \xi^n \rangle$ is $3/((n+1)(n+3))$.}
\label{sphericcase}
\begin{tabular*}{\textwidth}{@{\extracolsep{\fill}}lrrrrrr|rrrrl@{}}
\hline
$\langle  \xi^n \rangle$ & & \multicolumn{1}{c}\textrm{NRQCD} & \multicolumn{1}{c}\textrm{QCDSR} & \multicolumn{1}{c}\textrm{DSE}  & \multicolumn{1}{c}\textrm{BLFQ} &  \multicolumn{1}{c}\textrm{CST} & & \multicolumn{1}{c}\textrm{DSE}    & \multicolumn{1}{c}\textrm{BLFQ} &  \multicolumn{1}{c}\textrm{CST} &
  \multicolumn{1}{c}\textrm{pQCD}  \\
\hline
 $\langle  \xi^2 \rangle$  &\multirow{5}{*}{$\eta_c$}    & 0.075(11)  & 0.070(7)  & 0.10     &0.12 & 0.16 & \multirow{5}{*}{$\eta_b$}     & 0.070     &0.071& 0.13 & 0.20 \\
 $\langle  \xi^4 \rangle$ & & 0.010(3) & 0.012(2)    & 0.032   & 0.036 & 0.061 & & 0.015      &0.015& 0.046 & 0.086 \\
 $\langle  \xi^6 \rangle$ & & 0.0017(7) & 0.0031(8)    & 0.015    & 0.014  &0.032 &&0.0042      &0.0051& 0.0232 & 0.047\\
 $\langle  \xi^8 \rangle$     &&    &   &    0.0059& 0.0068&0.0193&&    0.0012&0.0021& 0.0140  & 0.030 \\
   $\mu$ &   &$m_c$   &    $m_c$          &2 GeV & 1.7$m_c$  & 2$m_c$ & & 2 GeV       & 1.6$m_b$ &2$m_b$  &   \\
  \hline
 $\langle  \xi^2 \rangle$  &\multirow{5}{*}{$\eta'_c$}    &0.22(14)  & 0.18$^{+0.005}_{-0.07}$  &     &0.18 & 0.22 & \multirow{5}{*}{$\eta'_b$}     &     &0.10& 0.16 & 0.20 \\
 $\langle  \xi^4 \rangle$ & & 0.085(110) &  0.051$^{+0.031}_{-0.0031}$    &    & 0.059& 0.092  & &       &0.022& 0.056 & 0.086 \\
 $\langle  \xi^6 \rangle$ & & 0.039(77) & 0.017$^{+0.016}_{-0.014}$    &    & 0.025& 0.048&&      &0.0068& 0.0276 & 0.047\\
 $\langle  \xi^8 \rangle$     &&    &   &  & 0.012&0.029&&    &0.0027& 0.0165& 0.030 \\
   $\mu$ &   &$m_c$   &    $m_c$          &2 GeV & 1.7$m_c$  & 2$m_c$ & & 2 GeV       & 1.6$m_b$ &2$m_b$  &   \\
 \hline
 $\langle  \xi^2 \rangle$  &\multirow{5}{*}{$J/\psi^{||}$}    & 0.075(11)  & 0.070(7) & 0.039 &0.11& 0.15& \multirow{5}{*}{$\Upsilon^{||}$}     &     0.014 &0.061& 0.110& 0.20 \\
 $\langle  \xi^4 \rangle$ & &   0.010(3) & 0.012(2)  &    0.0038&0.030& 0.054    & &$4.3\times 10^{-4}$&0.012& 0.037& 0.086 \\
 $\langle  \xi^6 \rangle$ & &     0.0017(7) & 0.0031(8)    &    $7.3\times 10^{-4}$& 0.011& 0.027    &&    $4.4\times 10^{-5}$&0.0036& 0.0186& 0.047\\
 $\langle  \xi^8 \rangle$ & &    &   &    $3.3\times 10^{-4}$&0.0053& 0.0164    &&    $3.7\times 10^{-6}$&0.0014& 0.0112& 0.030 \\
   $\mu$ &   &$m_c$   &    $m_c$          &2 GeV & 1.7$m_c$  & 2$m_c$ & & 2 GeV       & 1.6$m_b$ &2$m_b$  &   \\
 \hline
  $\langle  \xi^2 \rangle$  &\multirow{5}{*}{$\psi(2S)^{||}$}    &   0.22(14)  & 0.18$^{+0.005}_{-0.07}$ &    &0.17& 0.20& \multirow{5}{*}{$\Upsilon'^{||}$}     &     &0.090& 0.131& 0.20 \\
 $\langle  \xi^4 \rangle$ & &       0.085(110) &  0.051$^{+0.031}_{-0.0031}$&    &0.053& 0.079    & &    &0.018& 0.041& 0.086 \\
 $\langle  \xi^6 \rangle$ & &      0.039(77) & 0.017$^{+0.016}_{-0.014}$&     &0.022& 0.040   &&    &0.0053& 0.0276& 0.047\\
 $\langle  \xi^8 \rangle$ & &    &   &    &0.010& 0.023     &&    &0.0020& 0.0111& 0.030 \\
   $\mu$ &   &$m_c$   &    $m_c$          &2 GeV & 1.7$m_c$  & 2$m_c$ & & 2 GeV       & 1.6$m_b$ &2$m_b$  &   \\
\hline
\label{tab:moments_DA}
\end{tabular*}
\end{table*}

\begin{table}
\centering
\caption{Rms relative velocity of valence constituents from DSE \cite{DSE2016}, BLFQ and CST results.}
\begin{tabular*}{\columnwidth}{@{\extracolsep{\fill}}c|  lll | lll | lll@{}}
\hline
\multirow{2}{*}{} &
      \multicolumn{3}{c}{$\langle v^2\rangle$ } &
       \multicolumn{3}{c}{$\langle v^4\rangle$ }&
        \multicolumn{3}{c}{$\langle v^6\rangle$ } \\
    &DSE& BLFQ & CST &  DSE & BLFQ & CST & DSE & BLFQ & CST     \\
    \hline 
     $\eta_c$ & 0.31 &0.36  &0.48&0.16  &0.18  & 0.31&0.11&0.10&0.22\\
    $\eta'_c$ &  &0.54  &0.67&  &0.30  & 0.46&&0.18&0.34\\
    $J/\psi^{||}$& 0.12 &0.33  &0.45&0.02  &0.15  & 0.27&0.01&0.08&0.19\\
    $\psi(2S)^{||}$&  &0.51  &0.61&  &0.27 & 0.39&&0.15&0.21  \\
        \hline
    $\eta_b$ & 0.21 &0.21  &0.39&0.07  &0.07  & 0.23&0.03&0.04&0.16\\
    $\eta'_b$ &  &0.30  &0.47&  &0.11  & 0.28&&0.05&0.19  \\
    $\Upsilon^{||}$& 0.04 &0.18  &0.33&0.00  &0.06 & 0.19&0.00&0.03&0.13\\
    $\Upsilon'^{||}$&  &0.27 &0.39&  &0.09  & 0.21&&0.04&0.19 \\
    \hline
\end{tabular*}
\label{tab:rmsvelocity}
\end{table}

In Fig.\,\ref{fig:DA_jpsi}, the longitudinal distribution amplitude $\phi^{||}(x)$ of $J/\psi$ is shown for CST and BLFQ, together with the pQCD asymptotic limit given by $6x(1-x)$, and the model of Swarnkar {\it et al.} \cite{Swarnkar}, labeled here as AdS/QCD${}^{\text{(ii)}}$. For completeness, we also present the longitudinal PDA of the pion calculated from the light-front holographic model of Brodsky and de T\'eramond \cite{Lepage1980}, labeled as  AdS/QCD${}^{(\text i)}$, and in which the DA equals $(8/\pi)\sqrt{x(1-x)}$.

The CST and BLFQ PDAs are narrower than the latter PDA for the pion, but visibly broader than the prediction from AdS/QCD${}^{\text{(ii)}}$ for the corresponding $J/\psi$ state. A significant difference is the fact than in this approach, contrary to CST and BLFQ,  the one-gluon exchange interaction is absent. This interaction modifies the short range behavior, to which the PDAs are particularly sensitive to.

\begin{figure*}
 \centering 
 \includegraphics[width=0.42\textwidth]{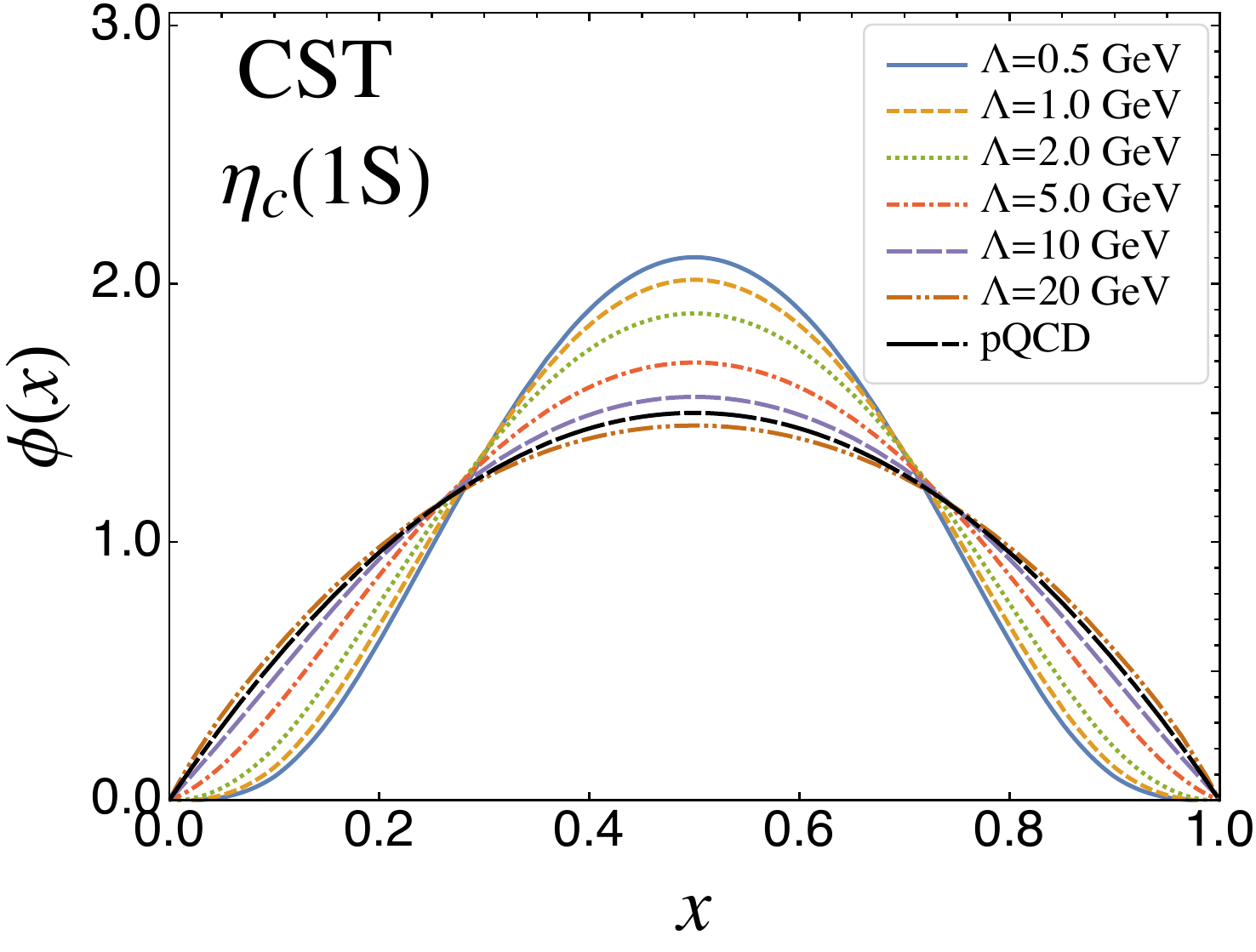}
 \includegraphics[width=0.42\textwidth]{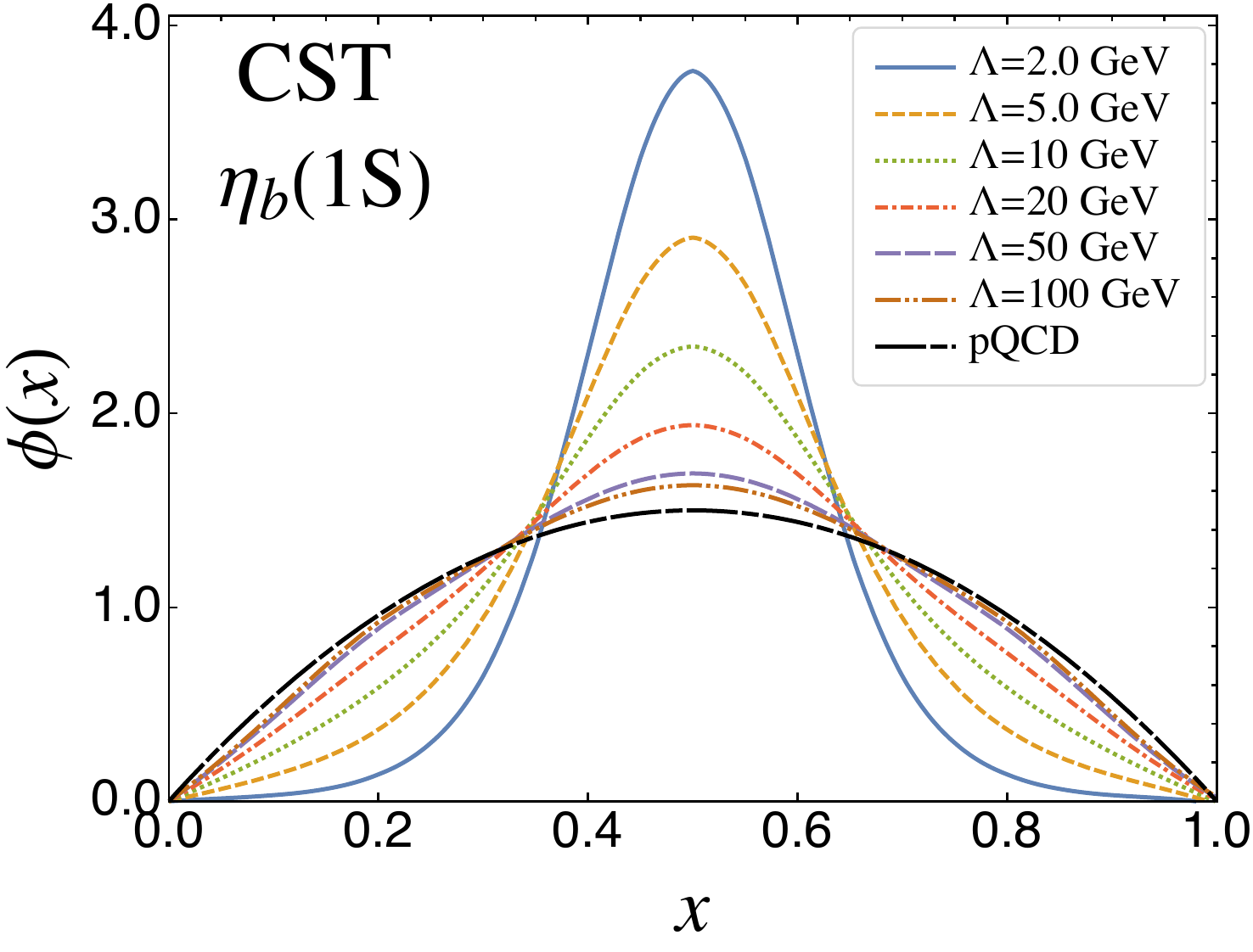}
  \includegraphics[width=0.42\textwidth]{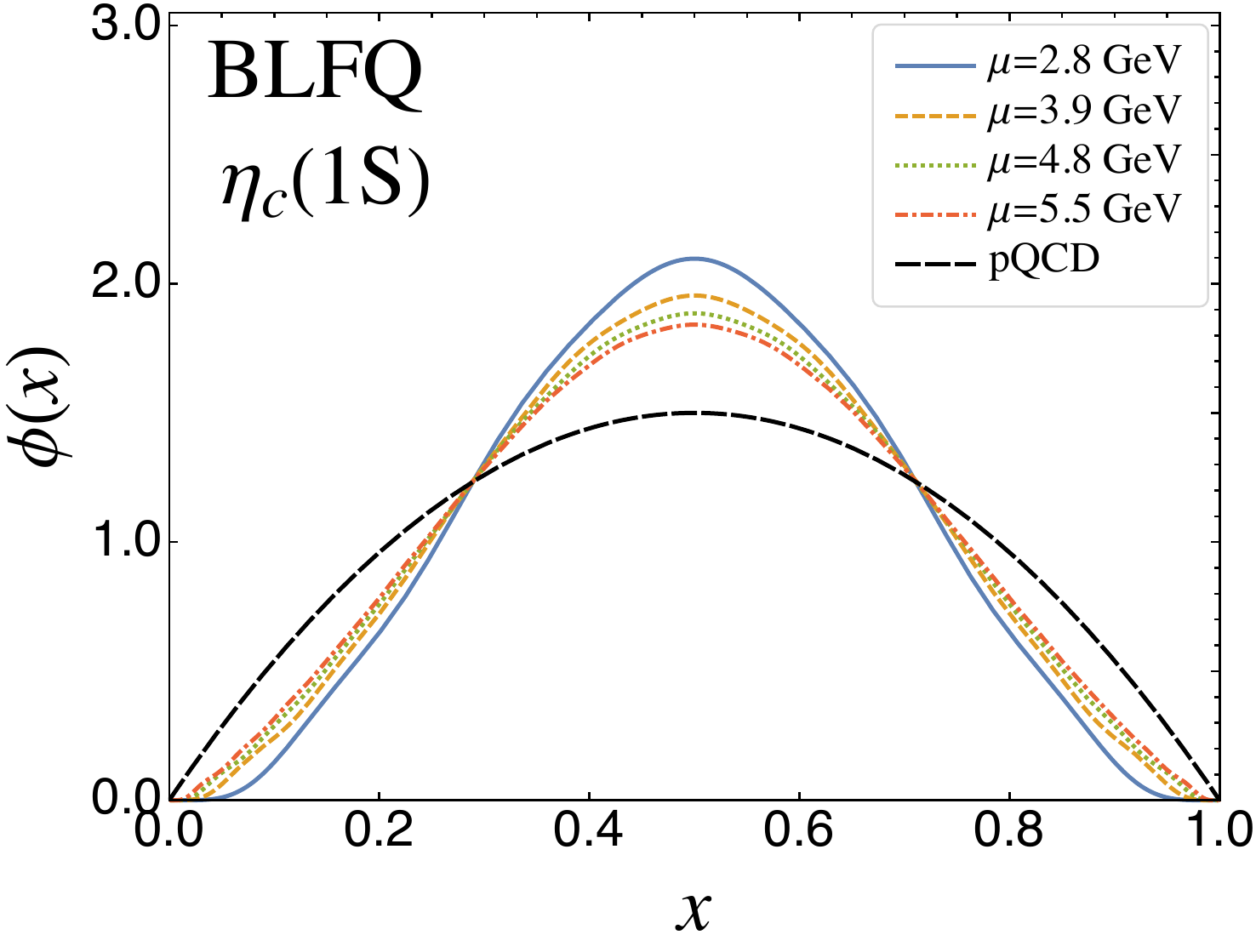}
 \includegraphics[width=0.42\textwidth]{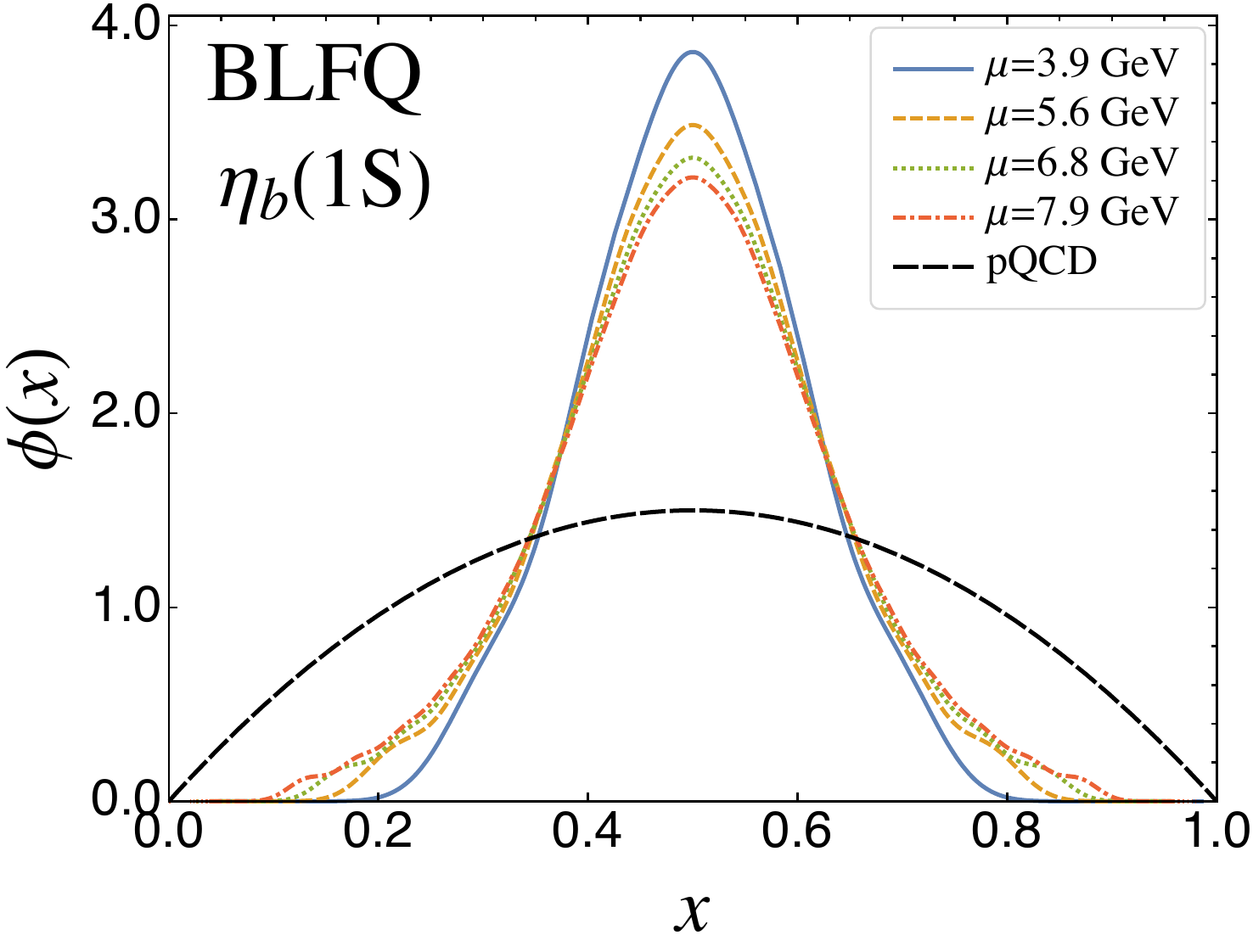}
 \caption{CST DAs calculated for increasing values of the cut-off parameter $\Lambda$ for $\eta_c$(1S) (top left) and $\eta_b$(1S) (top right); BLFQ DAs calculated for increasing values of the cut-off scale $\mu=\kappa\sqrt{N_{max}}$ (up to $N_{max}=32$)  for $\eta_c$(1S) (bottom left) and $\eta_b$(1S) (bottom right). The black dashed curve represents the pQCD limit  \cite{Lepage1980}.}
 \label{fig:DA_cutoff_CST_BLFQ}
 \end{figure*}

The CST and BLFQ PDA curves also differ from the pQCD limit, in principle, valid only at $\mu\to\infty$. Nevertheless, it is reassuring  to confirm that in the  CST approach,  by increasing the cut-off parameter $\Lambda$, indeed the PDAs get a broader shape and smoothly approximate the asymptotic perturbative QCD limit, as shown in Fig.~\ref{fig:DA_cutoff_CST_BLFQ} for both $\eta_c$ and $\eta_b$ states. In BLFQ, preliminary studies show that the pseudoscalar PDAs also approach the pQCD asymptotics as the cut-off scale increases, although larger $\Lambda_{UV}$ calculations are needed to confirm this point (cf. Fig.~\ref{fig:DA_cutoff_CST_BLFQ}).

Also, connected to the scale dependence, it is interesting to note that approaches with a smaller cut-off, typically of the order of the constituent quark mass or even smaller (DSE results), lead naturally to distributions with lower moments, as shown in Table \ref{tab:moments_DA}. CST and BLFQ, both have a larger cut-off, roughly of $2m$ (cf. section \ref{sec:definitions}), resulting in larger and comparable moments.
  
\begin{figure*}
 \begin{center}
    \includegraphics[width=.35\textwidth]{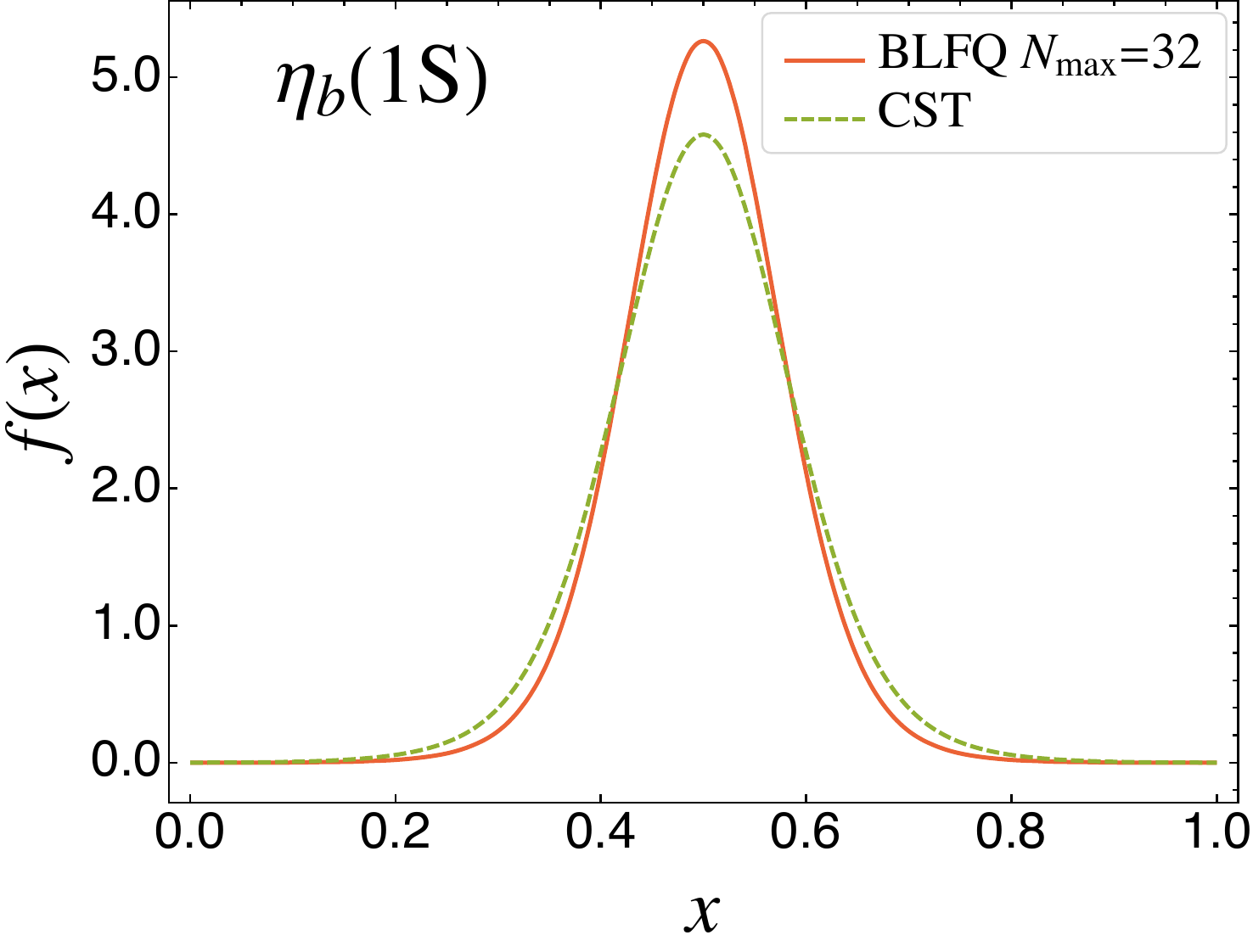}
  \includegraphics[width=.35\textwidth]{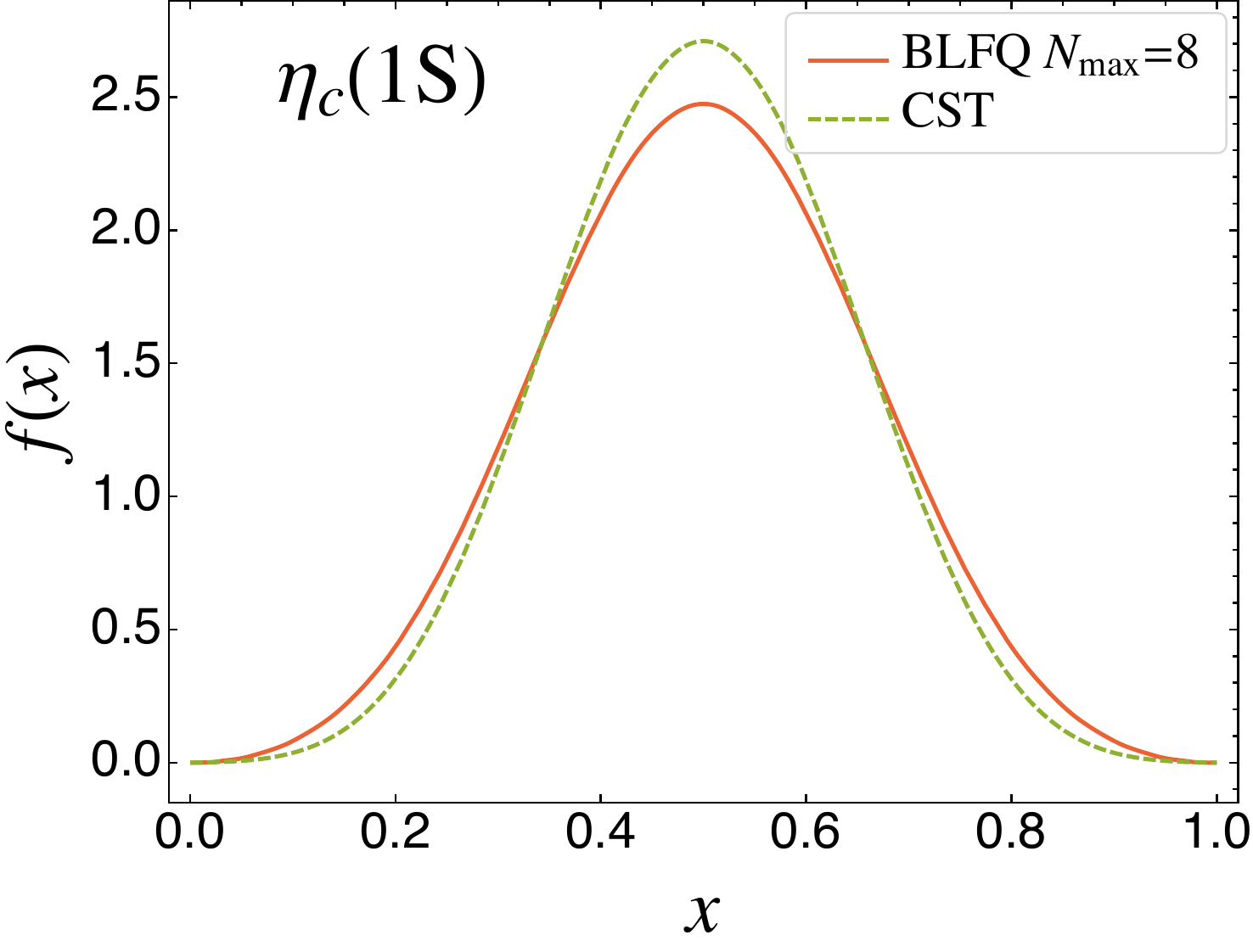}\\
      \includegraphics[width=.35\textwidth]{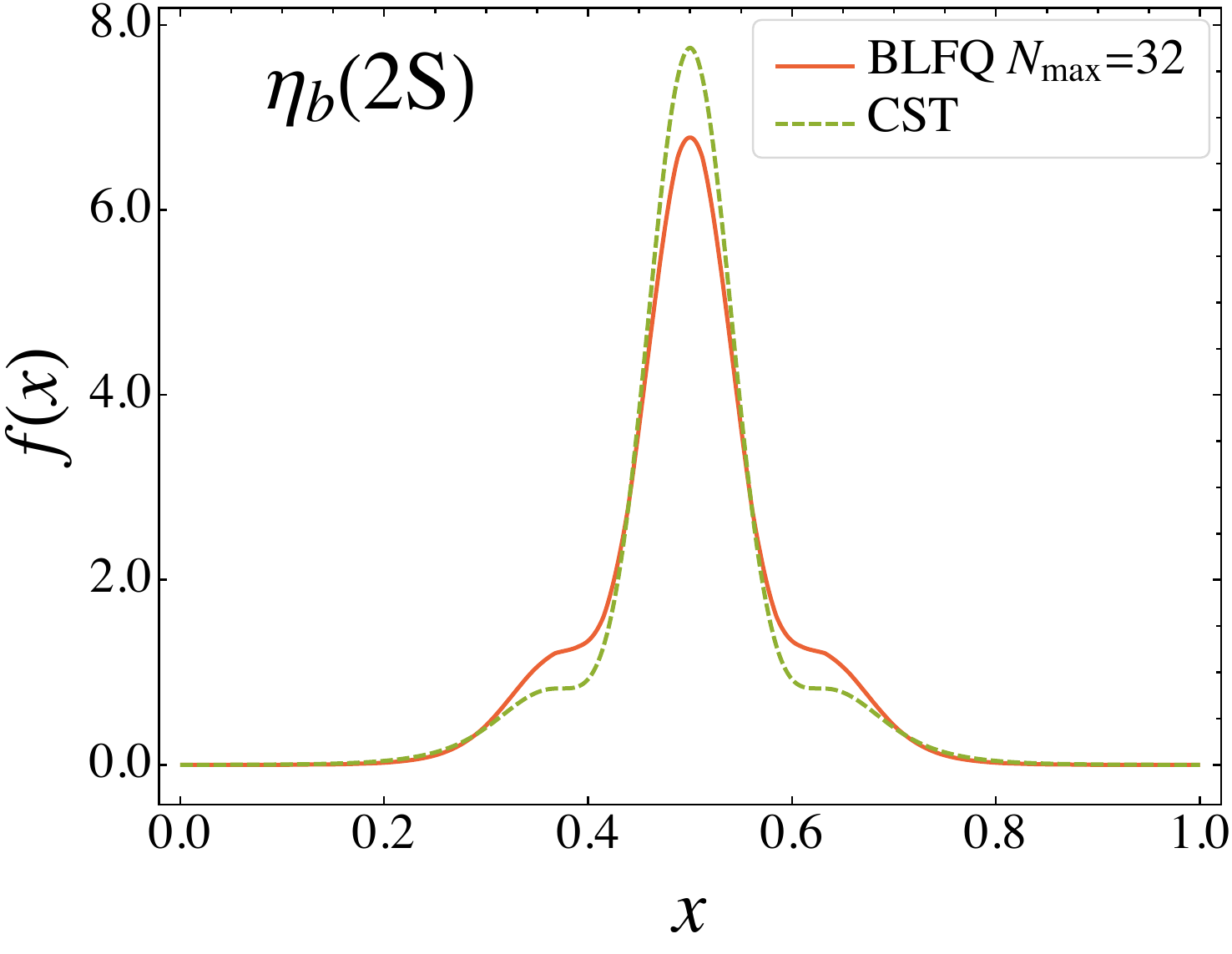}
  \includegraphics[width=.35\textwidth]{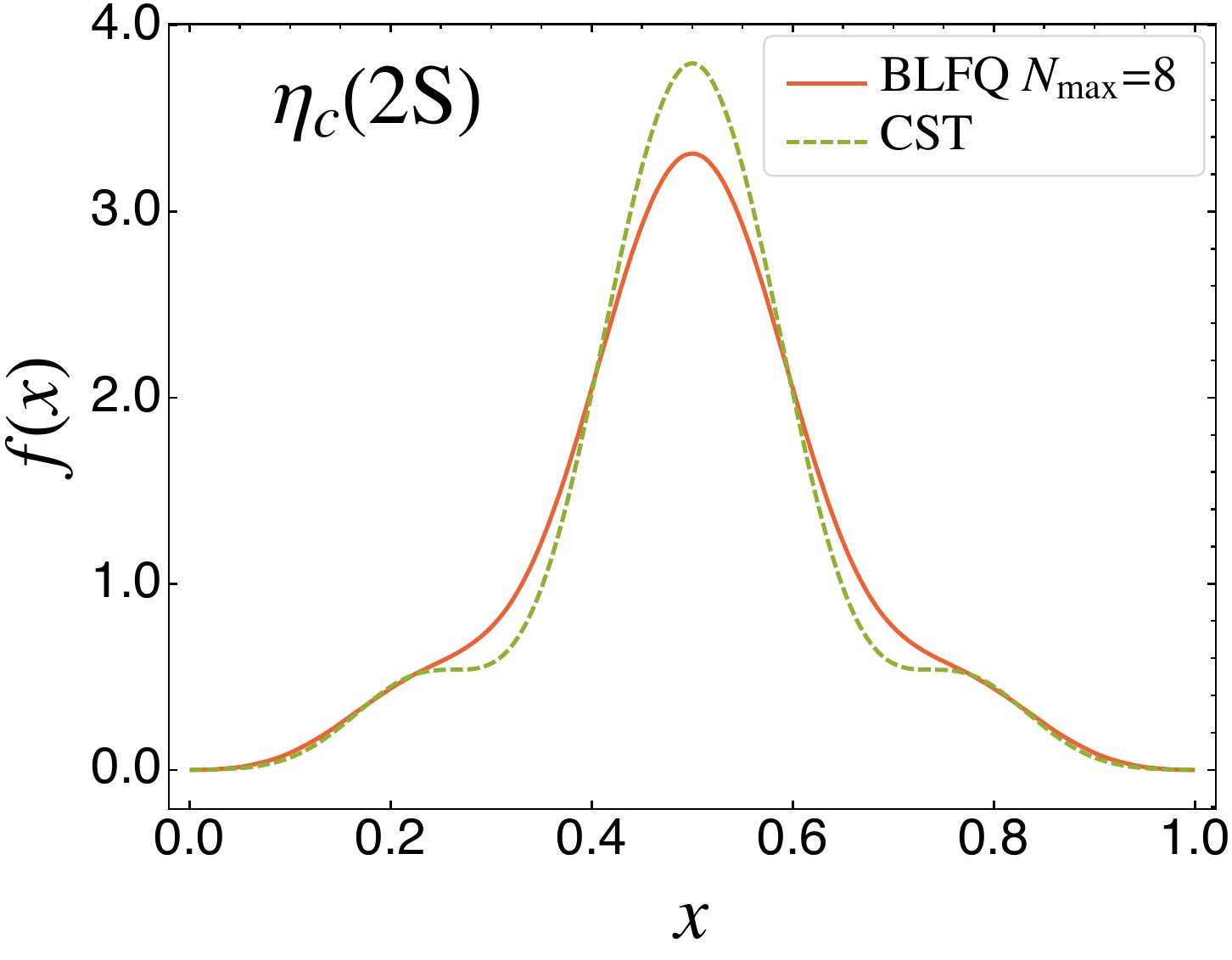}\\
  \caption{Parton distributions functions of pseudoscalar states.}
\label{fig:PDFs_pseudoscalar}
\end{center}
\end{figure*}

We also show the parton distribution functions (PDFs) in Fig.\,\ref{fig:PDFs_pseudoscalar}. Once again the results of the two approaches are consistent. In particular the first moments (displayed in Table \ref{tab:PDFmoments}) are very similar. On the other hand, the CST PDFs in Fig.\,\ref{fig:PDFs_pseudoscalar} tend to display more pronounced wavy structures than the BLFQ PDFs for the $\eta_b(2S)$ and $\eta_c(2S)$. 

\begin{table}
\centering
\caption{Parton distribution moments $\langle x^n\rangle$ for pseudoscalar states from (i) BLFQ and (ii) CST results.}
\begin{tabular*}{\columnwidth}{@{\extracolsep{\fill}}cllllll@{}}
\hline
 \multirow{2}{*}{PDFs} &
      \multicolumn{2}{c}{$\langle x^2\rangle$ } &
       \multicolumn{2}{c}{$\langle x^3\rangle$ }&
        \multicolumn{2}{c}{$\langle x^4\rangle$ } \\
    & (i) & (ii) &  (i) & (ii) &  (i) & (ii)   \\
        \hline
    $\eta_c(1S)$ & 0.274 & 0.270 & 0.161 & 0.155 & 0.1000 & 0.0937 \\
    $\eta_c(2S)$ & 0.271 & 0.269 & 0.157 & 0.153 & 0.0955 & 0.0919 \\
    \hline
        $\eta_b(1S)$ & 0.256 & 0.259 &  0.135 & 0.138 & 0.0724 & 0.0761\\
    $\eta_b(2S)$ & 0.257 & 0.257 & 0.136 & 0.135 & 0.0736 & 0.0728 \\
    \hline
\end{tabular*}
\label{tab:PDFmoments}
\end{table}

\section{Summary and Conclusions}
\label{sec:conclusions}

In this work we explore similarities and differences of two relativistic models of quarkonium. The successful results, reported in \cite{Li:2015zda} and \cite{Leitao:2016bqq} were improved, as confirmed by a refined spectroscopy, and extended with a new calculation of decay constants. We find our results for observables to be in good agreement with each other and with several theoretical approaches including lattice methods.
Beyond the level of observables, we used a map that allowed us to compare CST amplitudes with light-front wave functions and we observed a remarkable agreement between them.  We used these light-front wave functions to calculate parton distribution amplitudes and parton distribution functions. Our results again appear consistent  with each other. 

Noticeably, a first general conclusion is that in both approaches fixing the models solely to the mass spectra is sufficient to guarantee a reasonable overall description of the decay constants, whose precise description is known to be particularly challenging.

Since both approaches are  established in Minkowski space, we could apply and test the BHL prescription, allowing us to perform not only a benchmark comparison between the two obtained LFWFs, but also to have direct access to quantities such as light-cone distributions. These are extracted here in  a straightforward way, which is not possible in approaches relying on Euclidean formulations of quantum field theories.
 
 The combined analysis of the two approaches enriches our understanding and the predictive power of each model alone, providing robustness tests to each other. For instance, while in BLFQ the angular momentum is not a good quantum number, the agreement with CST suggests that the extraction of the angular momentum of each state is indeed reliable. Also, for the present 1CSE solutions, that do not possess a definite charge conjugation parity, the comparison provides a way to judge the deviations in observables due to that violation.
 
All these conclusions point towards the need of comparisons between different approaches like the one in this paper. This way one obtains control on model dependencies and isolates method-independent features in the spectrum and production processes. This may be an advantage in the future, for instance when  investigating the existence of exotic mesons.

Despite the formal difficulties, bottom up approaches shed light on difficult questions, such as how one may  bridge different approaches and combine knowledge.
This research has raised many questions, some of them in need of further investigation. One of them is certainly  whether or not analogous linkages could be developed for  lighter systems, which is planned for a future work.

\begin{acknowledgements}
This work was supported in part by the Portuguese Funda\c{c}\~ao para a Ci\^encia e Tecnologia (FCT) under contracts SFRH/ BD/ 92637/2013, SFRH/ BPD/100578/2014, and UID/FIS/0777/2013 and in part by the Department of Energy under Grant Nos. DE-FG02-87ER40371 and DESC0008485 (SciDAC-3/NUCLEI). Computational resources were provided by the National Energy Research Supercomputer Center (NERSC), which is supported by the Office of Science of the U.S. Department of Energy under Contract No.  DE- AC02- 05CH11231. S.L. thanks the hospitality of the Nuclear Theory Group of Iowa State University where most of this work was developed.
\end{acknowledgements}

\end{document}